\documentclass[10pt,a4paper]{article}
\usepackage[utf8]{inputenc}
\usepackage{float}
\usepackage{amsmath}
\usepackage{amssymb}
\usepackage{graphicx}
\usepackage{esint}

\makeatletter

\usepackage{cite}
\usepackage{caption}

\title{\sf\bf Analytical approach to high harmonics spectrum in the nanobunching regime}
\author{ Mykyta Cherednychek and Alexander Pukhov \\ \normalsize{\textsl{Institut f\"ur theoretische Physik, Heinrich-Heine-Universit\"at D\"usseldorf}}}

\makeatother

\begin{document}
\maketitle
\begin{abstract}
With the high-order harmonic generation (HHG) from plasma surfaces
it is possible to turn a laser pulse into a train of attosecond
or even zeptosecond pulses in the backward radiation. These attosecond
pulses may have amplitude several orders of magnitude higher than
that of the laser pulse under appropriate conditions. We study this
process in detail, especially the nanobunching of the plasma electron
density. We derive an analytical expression that describes the electron
density profile and obtain a good agreement with particle-in-cell
simulation results. We investigate the most efficient case of HHG at moderate
laser intensity ($I\approx2\cdot10^{20}W/cm^{2}$) on the over-dense plasma slab with an
exponential profile pre-plasma. Subsequently we calculate the spectra
of a single attosecond pulse from the backward radiation using our expression
for the density shape in combination with the equation for the spectrum of
the nanobunch radiation. 
\end{abstract}

\section{Introduction}

Since the invention of laser in the year of 1960 \cite{Maiman},
laser technology has witnessed an immense progress \cite{SM85,PSMH,PSBMH,PP,MU,WBCD,PM94,Mourou97,BUGK,NKZ,BRPC,MTB,Yanovsky,MT11,MMKS}.
The revolutionary invention of the chirped pulse amplification technique
brought the laser to a completely new level \cite{SM85}. Roughly
twenty years later a record peak intensity of order $10^{22}W/cm^{2}$
was reached by focusing a 45-TW laser beam \cite{BRPC}. Recently,
a compression scheme has been proposed that opens the possibility
to generate ultra-short and ultra-strong laser pulses with focused
intensities of $10^{24}W/cm^{2}$ and duration of 2fs \cite{MMKS}.

This progress offers an opportunity to study new physical phenomena
of laser plasma interactions. One of the most important processes in
this field is the HHG, which has been studying very intensely nowadays. As the minimum achievable duration
of laser pulses continuously reduces towards few femtoseconds,
the generation of even shorter pulses (in the attosecond or even zeptosecond
range) is possible only for radiation with shorter wavelengths. The
reduction of the pulse duration and the radiation wavelength would
open new horizons for potential applications. This is the main motivation
to study the HHG. 

First observations of HHG from plasmas were made in 1981 \cite{CFKa,CFKb}.
Rather matured is HHG in gases that allows to generate single attosecond pulses with duration less than 1fs \cite{HKS,BUU,SBC,GSH}.
However, this method of HHG requires the limitation of laser pulse
intensity by maximum $10^{15}W/cm^{2}$ in order to prevent the ionization.

Fortunately, there is another method of efficient production of high
order harmonics by unlimited laser power. This is the interaction
process of high contrast laser pulses \cite{TQG} with solid density
targets. The pedestal of the pulse ionizes the surface and the main
pulse interacts with electrons of the overdense plasma, while ions remain
nearly immobile during the short pulse duration. One distinguishes
two main HHG mechanisms in this case: coherent wake emission (CWE)
\cite{TQG,QTM,HHS} and the ``relativistically oscillating mirror'' (ROM)
\cite{BNP,LVP,LR,GPSB,GPSB2,BGP,DZG,DKB,HHM}. 

CWE is caused by fast Brunel electrons \cite{Brunel}, which excite plasma oscillations
at the local plasma frequencies. Thus, there are no harmonics beyond
the maximal plasma frequency in the case of CWE. This process dominates
for non-relativistic laser intensities $a_{0}\lesssim1$. For $a_{0}\gg1$
the harmonics are generated mostly via the ROM mechanism. In this case, the electron
layer at the plasma surface acts as a mirror that oscillates at relativistic
velocities, resulting in the generation of high order harmonics via Doppler effect when the surface moves towards
the incident wave. During this process there is no limit of frequency
like by CWE, so higher harmonics can be generated. The first theoretical
description of ROM claimed that the intensity spectrum envelope of
reflected wave can be described by $I(n)\propto n^{-5/2}$ up to the “roll
over'' frequency $\omega_{r}$ which is proportional to $4\gamma^{2}$, where
$n$ is the harmonic order and $\gamma$ is the relativistic gamma
factor \cite{GPSB}. Later this theory was improved, especially the
acceleration of the reflecting layer was taken into account. This
leads to the power law $I(n)\propto n^{-8/3}$ and $\omega_{r}\propto\gamma^{3}$
\cite{BGP}. This model assumes the existence of a so called apparent
reflection point (ARP) where the transverse electric field vanishes.
Predictions based on that model where experimentally confirmed \cite{DZG,DKB,HHM}. 

Most recently another HHG mechanism was discovered. Using p-polarized oblique incident
light with $a_{0}\gg1$ one can cause the formation of extremely dense
electron nanobunches under appropriate conditions. These bunches can
emit attosecond pulses with intensities much larger compared to the incident
pulse \cite{BP2,BP}. This means that the boundary condition assumed
in \cite{BGP} corresponding to ARP fails and thus the ROM theory
can not be applied in this case. This process is called coherent synchrotron
emission (CSE). The reflected radiation in case of CSE in characterized
by the power law $I(n)\propto n^{-4/3}$ or $I(n)\propto n^{-6/5}$
which is flatter comparing to ROM \cite{BP2,BP}. The corresponding
experiments can be found in Ref. \cite{DRY,DCR,YDC}. Detailed numerical
investigation of the case of p-polarized oblique incidence in Ref.
\cite{GKMS} demonstrate that the ROM model can be violated when the
similarity parameter $S=n/a_{0}$ (where $n$ is the electron density
given in units of the critical density $n_{c}$ and $a_{0}$ is the
dimensionless laser amplitude \cite{GP}) is smaller than five. The
authors of \cite{GKMS} present a new relativistic electronic spring
(RES) model for $S<5$.

Since usually one obtains a train of attosecond pulses by HHG, the
question is whether it is possible to isolate one single
pulse. One method is to use the polarization gate technique \cite{RGM,BaevaRelControl}. 
This is important because it opens the opportunity to a number of
potential applications \cite{PukhovNat}. Successful application of
$\lambda^{3}$ focusing could even lead to investigation of vacuum
instabilities \cite{NNS,GPSB2}.

We pursue two main goals in this work. The first one is to provide a more detailed analytical description of the spectrum in the case of CSE compared to \cite{BP,BP2}. 
For this purpose  we introduce an analytical approach which allows us to calculate the electron density profile of the given nanobunch as well as its current distribution, 
that are used in formulas for back-radiating spectrum derived in \cite{BP,BP2}. 
Subsequently we compare the derived expressions with one-dimensional simulation results done with the VLPL PIC code. 

The second aim is to determine the most efficient case of HHG at moderate laser intensity ($I\approx2\cdot10^{20}W/cm^{2}$). 
For that reason we perform several 1D PIC simulations with different parameters. Finally, we analyze the obtain results and define different regimes of HHG.

In the last section of the paper we consider the nanobunches moving and radiating in forward direction.

\section{PIC simulation of the HHG process}

\begin{figure}[htbp]
	\centering 
	\includegraphics[width=1\textwidth]{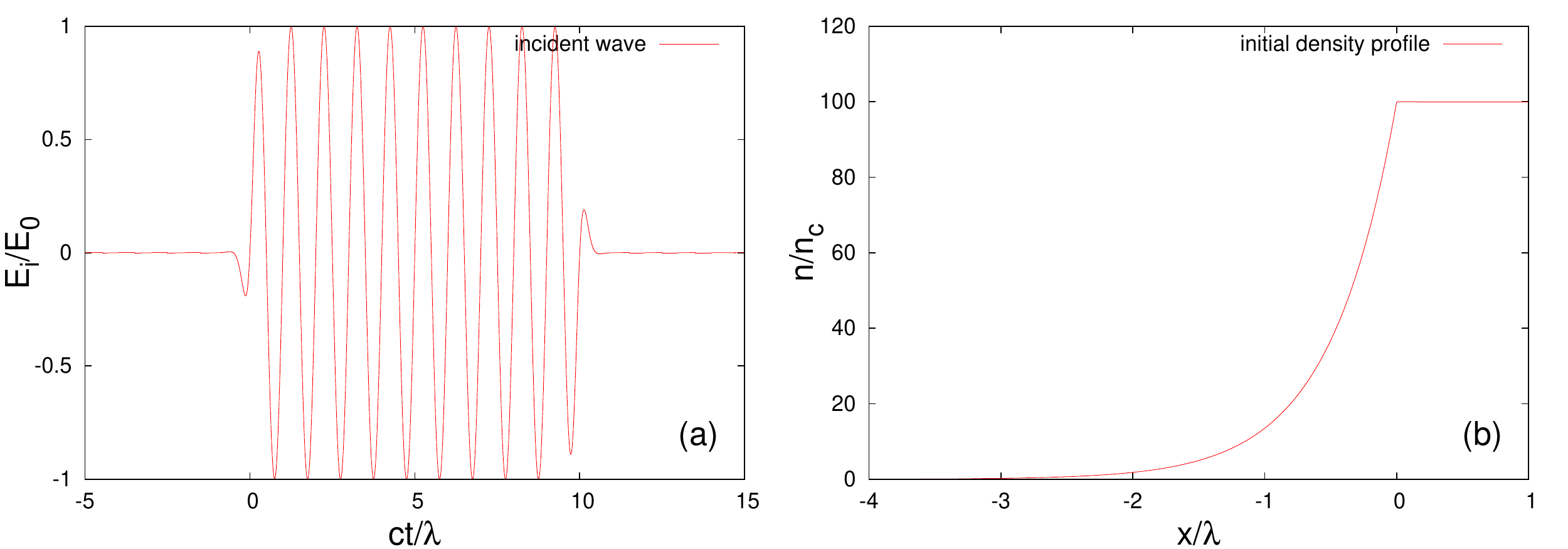} 
	\caption{\small (a) Electric field component of the incident electromagnetic
		wave in vacuum plotted versus time at $x=0$. (b) Initial density profile $\sigma=0.5\lambda$, $n_{0}=100n_{c}$,
		where $n_{c}$ is the critical density. }
	\label{Pulse}  
\end{figure}

For our simulations we use the one-dimensional version of the VLPL PIC code \cite{vlpl}. 
In our geometry, the incident wave comes from the left hand side of the simulation box
and propagates along the $x$-axis. The wave is p-polarized and the electric
field component oscillates along the $y$-axis. The plasma is located
at the right hand side of simulation box. It is also possible to simulate oblique incidence with our code.
Let $\theta$ be the angle of incidence in the laboratory frame and consider
a frame moving along the $y$-axis with velocity $V=c\sin\theta$.
Lorentz transformations verify that the laser is normally
incident in this frame. At the same time the whole
plasma moves in $y$-direction in the frame. Thus, attributing some
initial velocity to plasma in our simulation, we are working in the
moving frame. If we need to get the results in laboratory frame, we have
to transform the values obtained from the simulation via Lorenz transformation.
Consequently we obtain results that correspond to the process with
oblique incidence. We use the incident wave $E_{i}(\tau)$ of duration
$T=10\lambda/c$, that is given by 
\[
E_{i}(\tau)=\frac{1}{4}\left(1+\tanh\left(\frac{\tau}{\Delta t}\right)\right)\left(1-\tanh\left(\frac{\tau-T}{\Delta t}\right)\right)\sin(2\pi\tau),
\]
where $\Delta t=\lambda/4$ and $\tau=t-x/c$ (Fig. \ref{Pulse}a). 
Further we use an exponential plasma density ramp for $x<0$. For
$x>0$ the density remains constant (Fig. \ref{Pulse}b), 
\begin{align}
n(x)=\left\lbrace \begin{aligned} & n_{0}e^{\frac{x}{\sigma}}\qquad\text{for}\quad x<0\\
 & n_{0}\qquad\quad\,\text{for}\quad x>0
\end{aligned}
\right..\label{rump}\\
\nonumber 
\end{align}

\begin{figure}[htbp]
	\centering \includegraphics[width=1\textwidth]{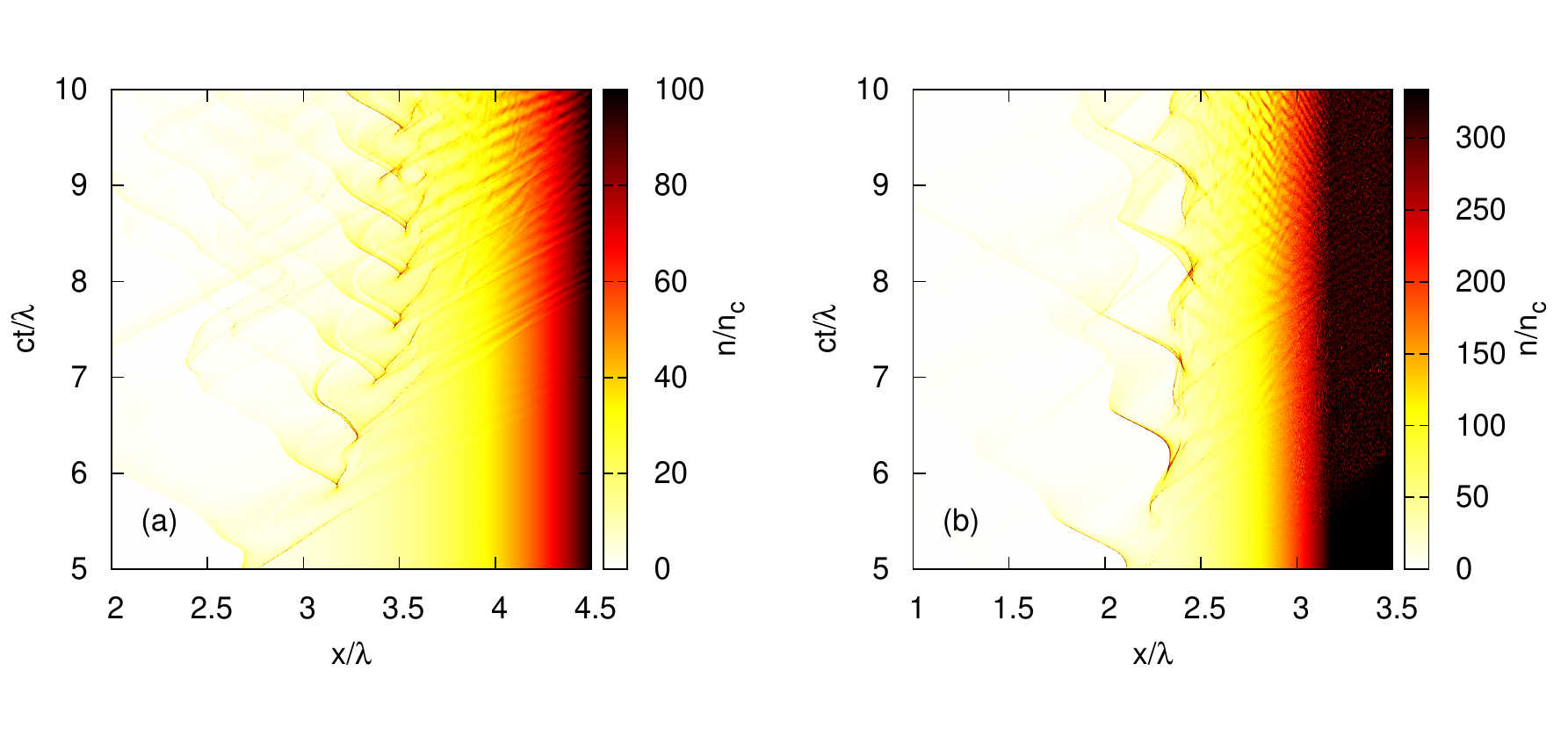} 
	\caption{\small The electron density distribution in space time domain
		by (a) normal incidence and (b) $48^{\circ}$ oblique incidence. Simulation
		parameters considering from the laboratory frame are: plasma density
		$n_{0}=100n_{c}$, $\sigma=0.5\lambda$; laser amplitude $a_{0}=10$.
		Note that the values in (b) are transformed concerning to the moving
		frame. }
	\label{dens_comp} 
\end{figure}

Assuming that the ions are at rest during the whole interaction process,
we consider only the interaction between the electrons and the incident
wave. In the simple case of normal incidence there are two forces
acting on particles along $x$-axis. The electrostatic force proportional
to $E_{x}$ and laser ponderomotive force oscillating with $2\omega$
(twice of the laser frequency). Thus the plasma surface oscillates
with the half of the laser period (Fig. \ref{dens_comp} (a)). 
In the case of oblique incidence of a p-polarized wave, there is going to be an additional
longitudinal component of the electric field oscillating at frequency
$\omega$ and acting on the plasma surface. Consequently the interaction becomes
even more complicated, which leads to stronger oscillations of the plasma
surface containing both $\omega$ and $2\omega$ modes (Fig. \ref{dens_comp}
(b)).

As soon as the electrons are pulled back by the electrostatic
force, they form a thin nanobunch that reaches a velocity close to $c$.
In this case the generation of high harmonics is possible.

\section{Density profile of a thin electron layer}
\label{ch:DensProf}

In this section we do the first step towards our first goal described in the introduction and derive two different analytic expressions for two
different cases, which roughly describe the electron density profile
at the times where the sharp spikes appear. The starting point of
our calculations is the approximation of the electron phase space
distribution at these times. As we shall see later this distribution
depends on the propagation velocity $\dot{x}_{0}(t)$ of the given electron
layer.

First, let us consider the case of a slow electron
bunch $\dot{x}_{0}(t)\ll c$. 
\begin{figure}[htbp]
	\centering \includegraphics[height=4.9999cm]{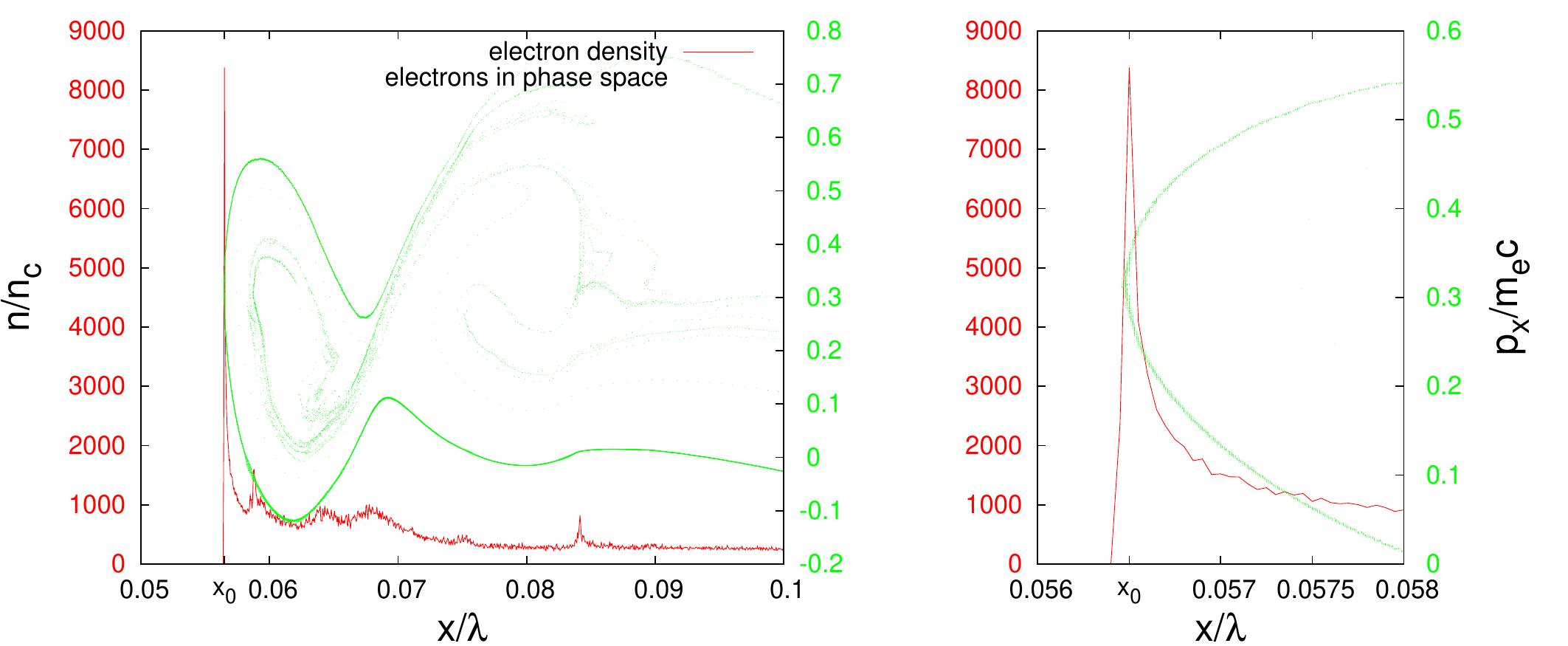} 
	\caption{\small Electron density (red) and electrons in $x$-$p_{x}$-plane
		(green). $x_{0}$ is the position of the maximal density. Simulation
		parameters: initial plasma density $n_{0}=241n_{c}$; $\sigma=0.001\lambda$,
		Pulse with dimensionless amplitude $a_{0}=10$ and p-polarized oblique
		incidence at $57^{\circ}$ angle has the wave length $\lambda=820$nm.
		All magnitudes are taken in the simulation frame. The right picture
		gives a zoom of the area around $x_{0}$.}
	\label{dens1_377} 
\end{figure}
In Fig. \ref{dens1_377} the electron density and its distribution
in $x$-$p_{x}$-phase space at time $t=0.875\lambda/c$, when the electrons are pushed inside the plasma almost to the maximal distance by the ponderomotive force  are visualized. 
We count the time according to Fig. \ref{Pulse} (Fig. \ref{Pulse}a shows the field oscillations at the point $x=0$ where the region with the constant plasma density begins (Fig. \ref{Pulse}b)).
Because we start with a cold plasma, the electron distribution function is
a (curved) line in the phase space. We assume that this curve in
phase space is described by the function $x(p)$ at some small interval
close to the density spike. Obviously, $x_{0}$ is the local minimum
of this function that coincides with the position of the spike. In
fact, we have always a spike of electron density at the point, where
the function $x(p)$ exhibits the local extreme value. For instance,
if we take a look at another curve in phase space, which is enclosed
by the previous one, we see that there is also a local maximum of
the density at the point were the curve reaches its minimal $x$-value.
However the electrons are more scattered (heated) compared to the
previous case and thus the local density maximum is much smaller.
The idea, that gives us the staring point for our calculations is
the following one. We can locally describe given curve in phase space
as a parabola: 
\begin{equation}
x(p,t)=x_{0}(t)+\alpha(t)(p-p_{0}(t))^{2}.\label{x(p)}
\end{equation}
The point ($x_{0}(t)$,~$p_{0}(t)$) corresponds to the local minimum.
We consider some short interval $\Delta x$ where this assumption
makes sense. Starting from this assumption and doing some algebra we obtain the expression for electron density profile 
\begin{equation}
n(x,t)=\frac{1}{2}\frac{N}{\sqrt{\Delta x\left(x-x_{0}(t)\right)}},\label{dens2}
\end{equation}
where $N$ is the number of particles contained between $x_{0}(t)$
and $x_{0}(t)+\Delta x$. Note that the parameter $\alpha$ cancels out,
so it does not affect the density profile.  
The derivation of this expression is shifted to the appendix \ref{ch:appendix}.
\begin{figure}[htbp]
	\centering \includegraphics[height=8cm]{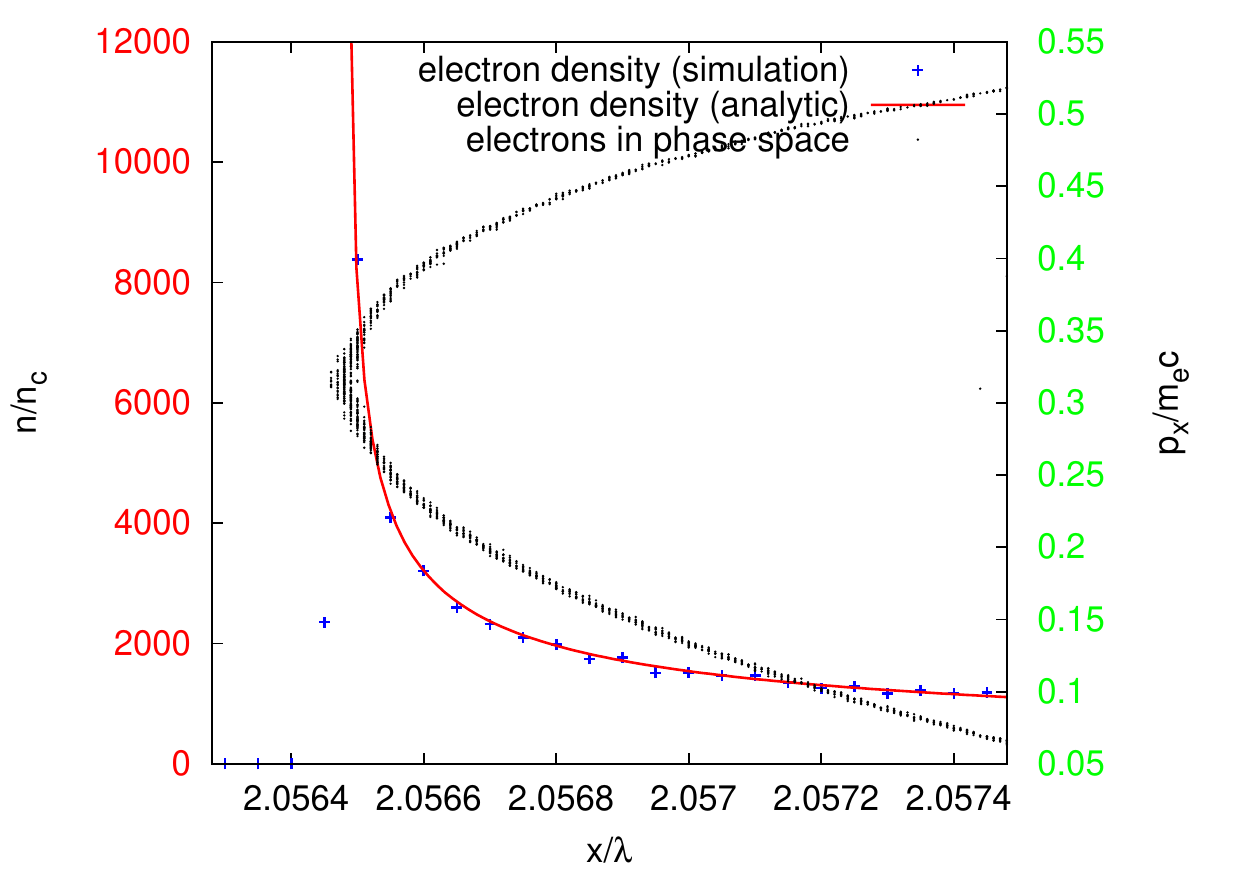} 
	\caption{\small Electron density taken from simulation (blue) and calculated
		analytically via (\ref{dens2}) (red), as well as electrons in $x$-$p_{x}$-plane
		(black), with same simulation parameters as for Fig. \ref{dens1_377},
		$\Delta x=0.001\lambda$ (simulation frame).}
	\label{fit_dens_377} 
\end{figure}
In Fig. \ref{fit_dens_377} we see that the density described with
(\ref{dens2}) agrees very well with simulation results. This picture
is actually a zoom of the Fig. \ref{dens1_377} at the position of
density spike. We obtain the best agreement at the instants when the
electrons are pushed inside at the maximal distance. In this case, the
mean momentum of the electrons is close to zero and equation (\ref{x(p)})
describes electrons in phase space quite well. We call the case where
$\dot{x}_{0}(t)\ll c$ is valid ``parabolic case''.

\begin{figure}[htbp]
	\centering \includegraphics[width=1\textwidth]{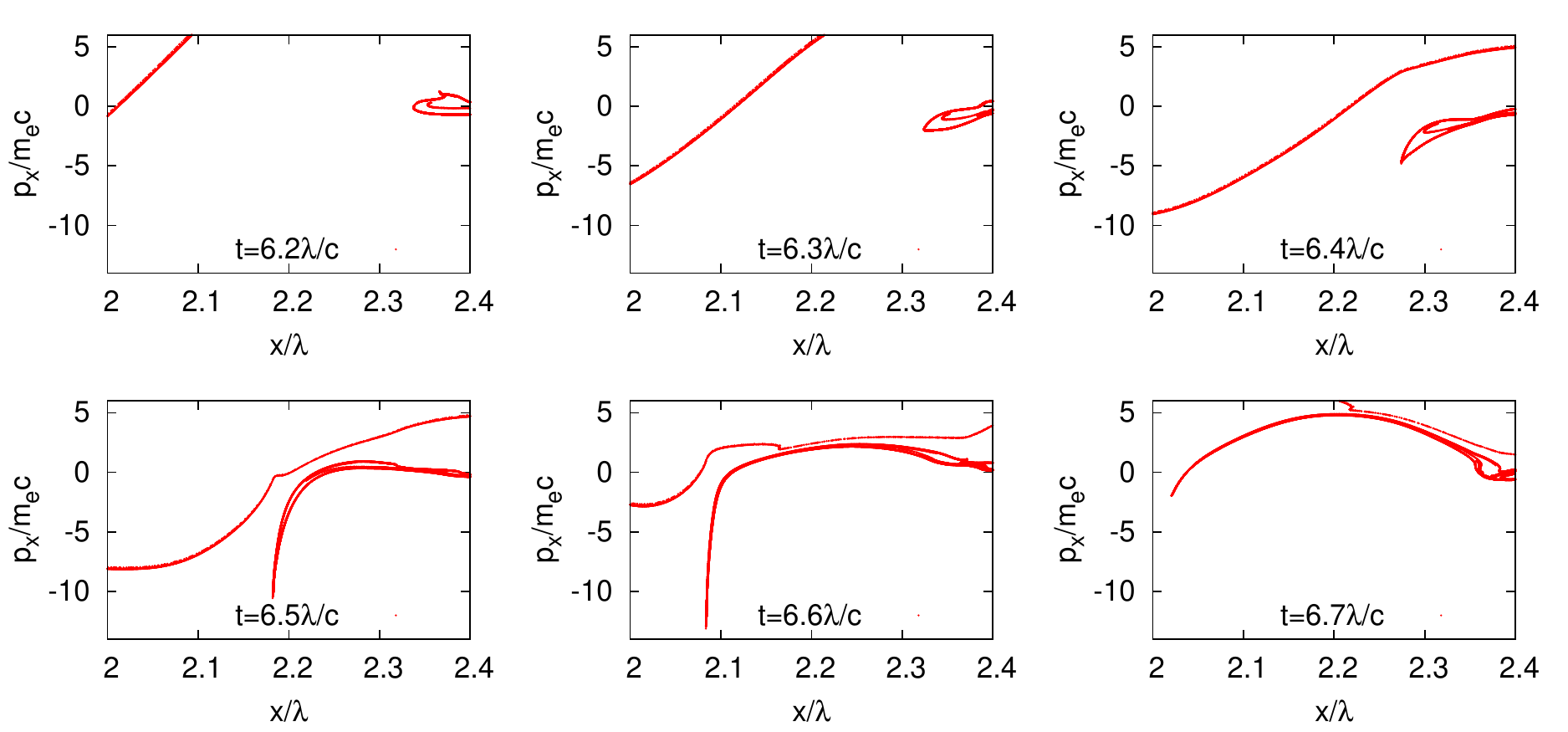} \caption{\small Electrons in $x$-$p_{x}$-plane taken from the simulation
		to different times $t$ during the process of nanobunching. Simulation
		parameters: initial plasma density $n_{0}=100n_{c}$; $\sigma=0.5\lambda$
		(laboratory frame), Pulse with dimensionless amplitude $a_{0}=10$
		and p-polarized oblique incidence at $48^{\circ}$ angle has the wave
		length $\lambda=820$nm.}
	\label{xpx2} 
\end{figure}

\begin{figure}[htbp]
	\centering \includegraphics[height=6cm]{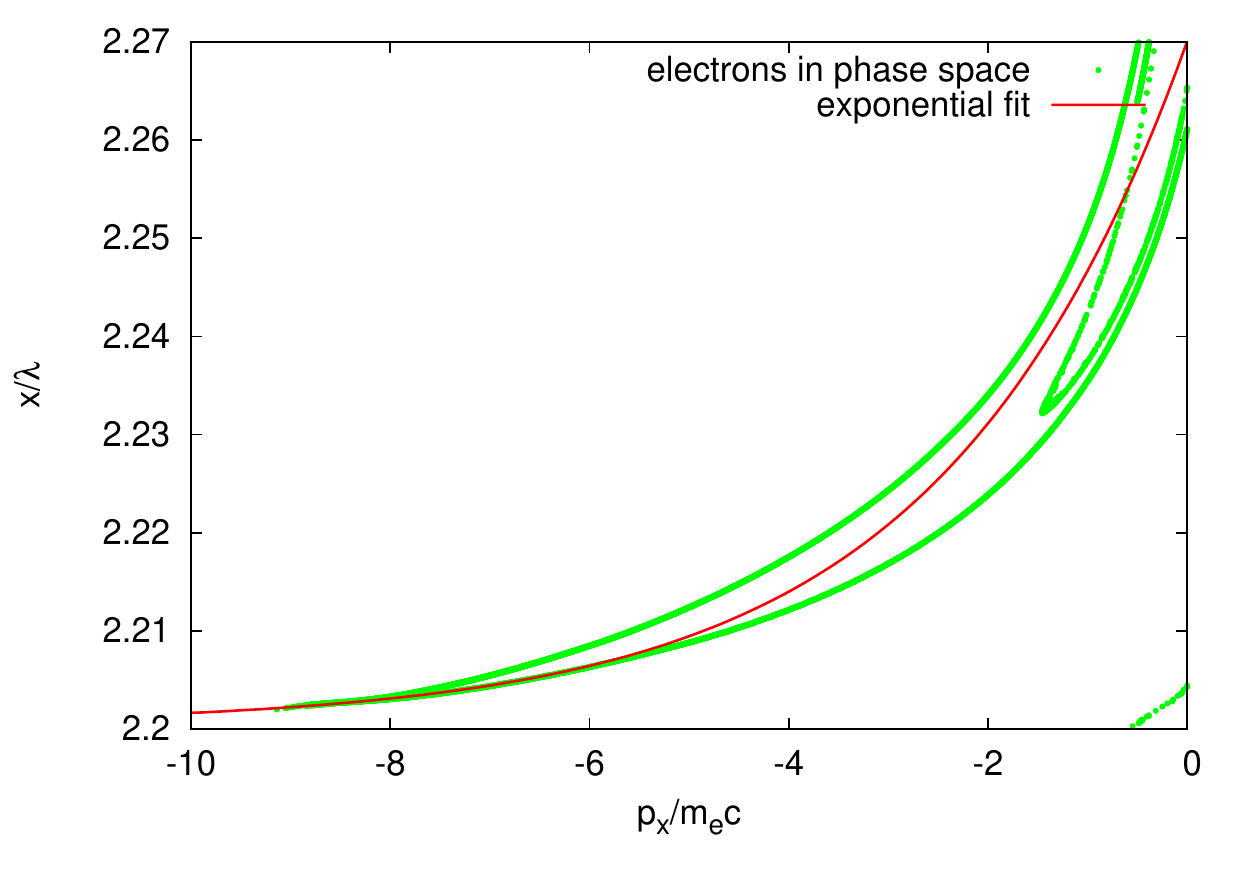} 
	\caption{\small Electrons in $x$-$p_{x}$-plane (green) and the exponential
		fit (red) from the same simulation as Fig. \ref{xpx2}, taken to the
		time $t=6.48\lambda/c$.}
	\label{xpx} 
\end{figure}

\begin{figure}[htb]
	\centering \includegraphics[width=1\textwidth]{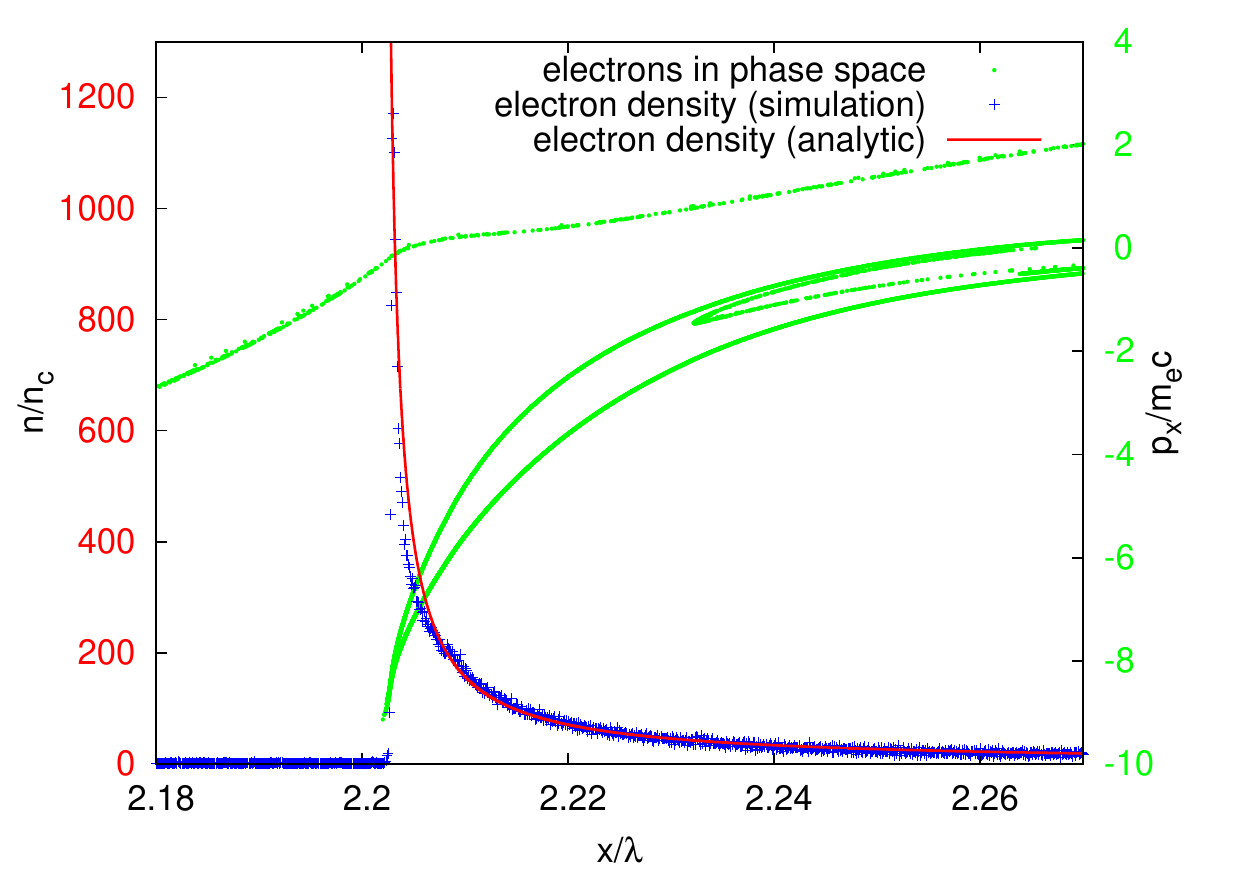} 
	\caption{\small Electron density taken from simulation (blue) and calculated
		analytically via (\ref{dens1/x2}) (red), as well as electrons in
		$x$-$p_{x}$-plane (green), with same simulation parameters compared
		to Fig. \ref{xpx2}, taken at $t=6.48\lambda/c$. $x_{\text{min}}-x_{0}=8,5\cdot10^{-4}\lambda$;
		$x_{\text{max}}-x_{0}=0.05\lambda$ (simulation frame). }
	\label{fit_dens_exp1}  
\end{figure}

Now we discuss another case with $\dot{x}_{0}(t)\rightarrow c$. Consider
the phase space evolution taken from the other simulation illustrated
in Fig. \ref{xpx2}. At the beginning, $t=6.2\lambda/c$, the momentum
is close to zero and the distribution is parabolic as expected. Further,
as soon as the electron bunch is pulled back by the electrostatic force,
the negative momentum of the bunch grows constantly with time and
the distribution changes its form until it becomes a kind of a ``whip''
between $t=6.5\lambda/c$ and $t=6.6\lambda/c$. The extremely dense
electron nanobunch reaches the velocity close to $c$ during this
period. 
This picture is taken from the same simulation as Fig. \ref{dens_comp}
(b). In this case the phase space distribution can be roughly fitted
with an exponential function 
\begin{equation}
x_{p}(p,t)=x_{0}(t)+e^{\alpha(t)\left(p-p_{0}(t)\right)}\label{exp2}
\end{equation}
(see Fig. \ref{xpx}). 
As we will show later, the incident angle and the density gradient that were
used here are optimal for producing the most intense attosecond pulse.
The phase space distribution belongs to the nanobunch that emits
this pulse. In order to simplify the notation we drop the time dependence
and set $p_{0}=x_{0}=0$. At this point we have
\begin{equation}
x_{p}(p)=e^{\alpha p}\label{exp}.
\end{equation}
On some short interval $[x_{\text{min}}:x_{\text{max}}]$ the density profile can be calculated to
\begin{equation}
n(x)=\frac{N}{\ln{\left(\frac{x_{\text{max}}}{x_{\text{min}}}\right)}x},\label{dens1/x2}
\end{equation}
where $N$ is the number of particles contained between $x_{\text{min}}$ and $x_{\text{max}}$.
A technical details are shifted to the appendix \ref{ch:appendix}.
As we can see from Fig. \ref{fit_dens_exp1} equation (\ref{dens1/x2})
approximates the density profile quite well even over the comparably long
interval. 
We call the case where $\dot{x}_{0}(t)\rightarrow c$ is valid ``whip
case''.

Although the functions (\ref{dens2}) and (\ref{dens1/x2}) work well
on the given intervals, we still have the problem that they both are not
continuous or even exhibit a singularity. Such behavior is obviously
not physical. In fact we are able to describe only a part of the spike
correctly. In order to solve this problem we need to find an expression
that would describe the whole spike. That means for instance
on interval $[x_{0}-\Delta x:x_{0}+\Delta x]$. In order to find such
function we have to replace the delta function, which is used for the definition of the distribution function in equations (\ref{phase_space1})
and (\ref{phase_space2}) by some limited function $\delta_{a}(x)$
with 
\begin{equation}
\lim_{a\rightarrow0}\delta_{a}(x)=\delta(x).\label{delta_conv}
\end{equation}
The parameter $a$ describes the width of $\delta_{a}$, which means
that $a>0$ is required. One of the possible definitions of $\delta_a$ leads to the solution 

\begin{align}
n_{a,\Delta x}(x)=\left\lbrace \begin{aligned} & \frac{N_{a,\Delta x}}{5a^{3}\sqrt{\Delta x}}\left(3a^{2}-2x^{2}+ax\right)\sqrt{x+a}\qquad\qquad\qquad\text{for}\quad x\in[-a,a]\\
 & \frac{N_{a,\Delta x}}{5a^{3}\sqrt{\Delta x}}\biggr(\left(3a^{2}-2x^{2}\right)\left(\sqrt{x+a}-\sqrt{x-a}\right)\\
 & \qquad\qquad\qquad\quad+ax\left(\sqrt{x+a}+\sqrt{x-a}\right)\biggr)\qquad\text{for}\quad x>a\\
 & \qquad0\qquad\qquad\qquad\qquad\qquad\qquad\qquad\qquad\qquad\text{for}\quad x<-a
\end{aligned}
\right.\nonumber \\
\label{profile_sqrt}
\end{align}
See appendix \ref{ch:appendix} for more details. 

\begin{figure}[htb]
	\centering \includegraphics[width=10cm]{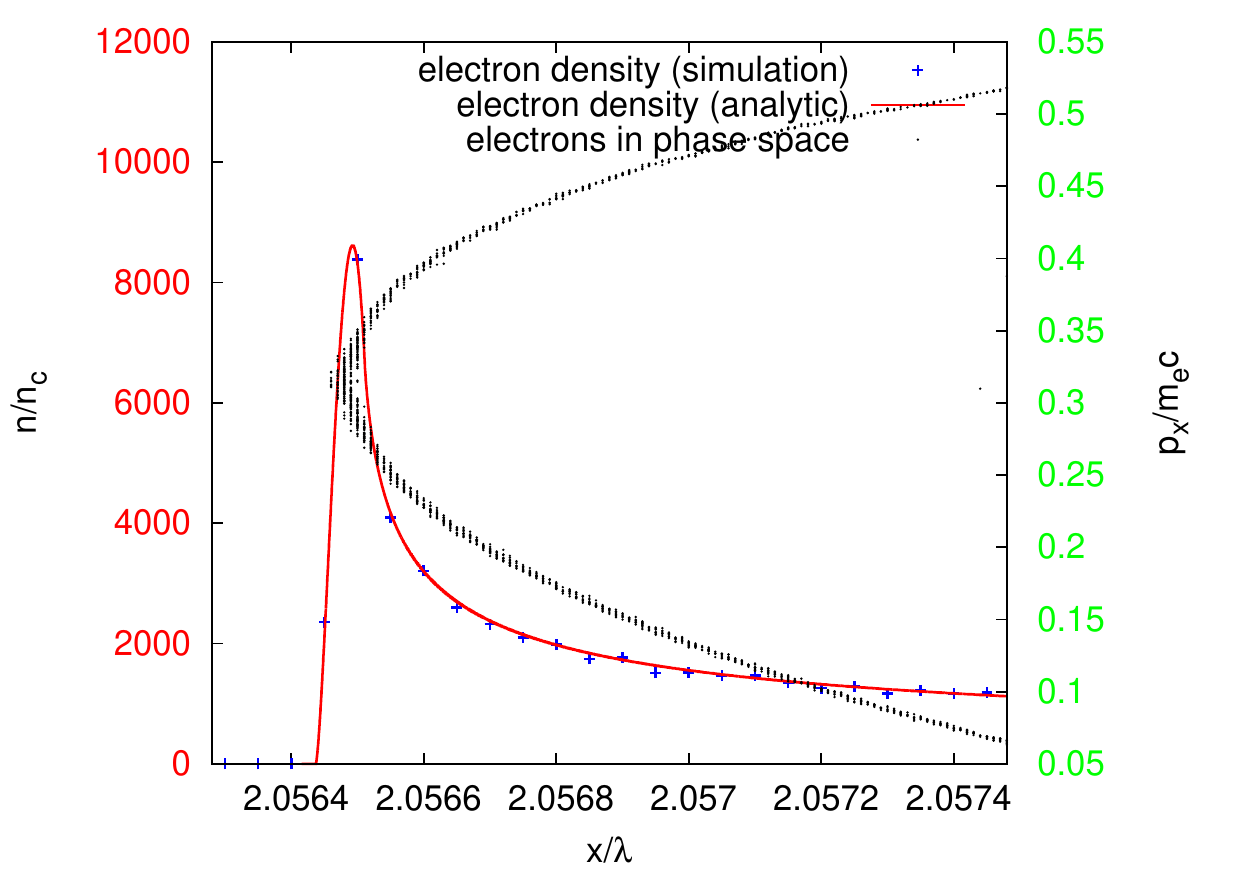} 
	\caption{\small Electron density taken from the simulation (blue) and
		calculated analytically via (\ref{profile_sqrt}) (red), with same
		simulation parameters as for Fig. \ref{dens1_377}, $\Delta x=0.001\lambda$;
		$a=3,7\cdot10^{-5}$ (simulation frame).}
	\label{plot_prof1} 
\end{figure}

Now let us try to fit the simulated density from above with the calculated
analytical profile (Fig. \ref{plot_prof1}). 
We chose a quite small value for $a$ because the plasma is cold and
we are dealing with very big and sharp spike as it is shown in this example. This
is the case since we use strong laser pulse and very small cell size
($5\cdot10^{-5}\lambda$). As we can see, our function agrees well
with the simulated profile.

\begin{figure}[htb]
	\centering \includegraphics[width=1\textwidth]{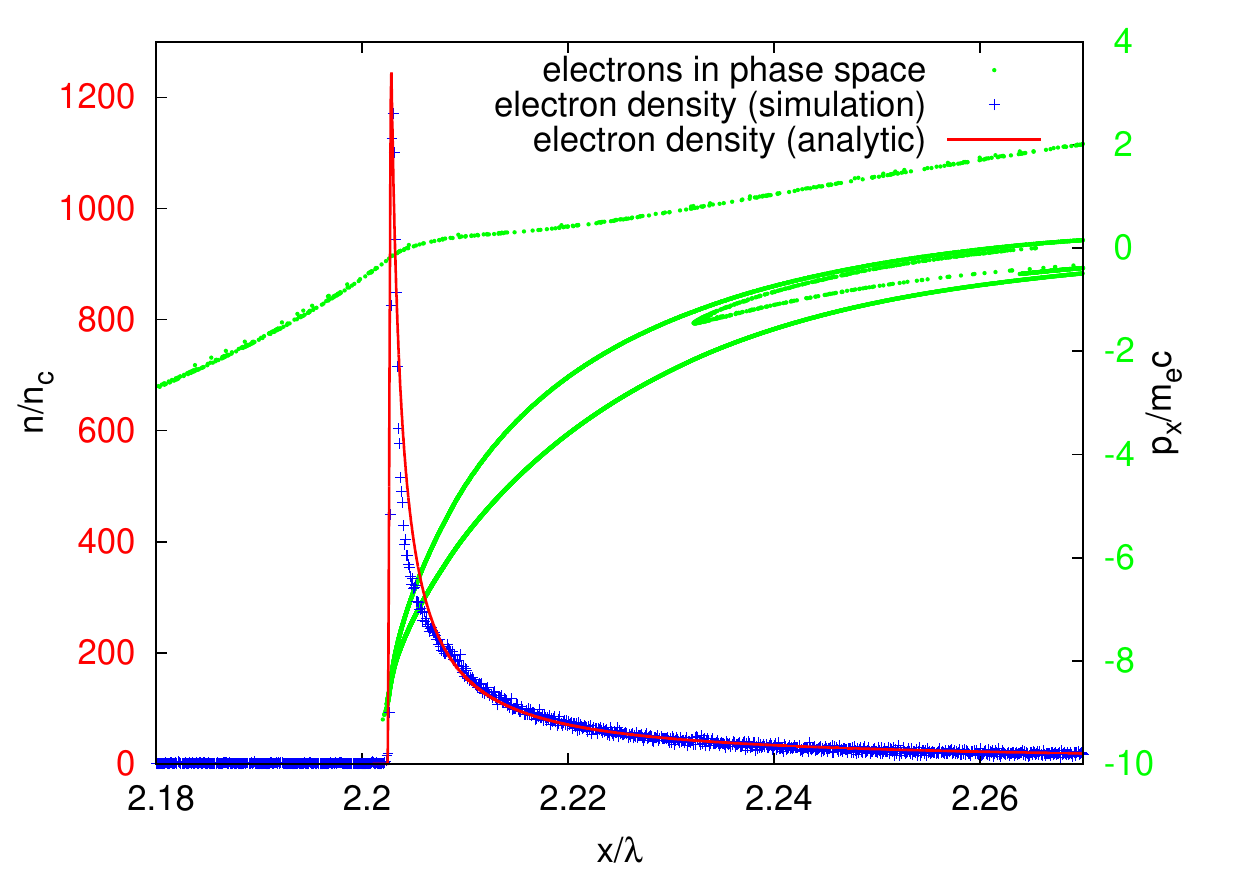} 
	\caption{\small Electron density taken from simulation (blue) and calculated
		analytically via (\ref{profile_exp}) (red), as well as electrons
		in $x$-$p_{x}$-plane (green), with same simulation parameters compared
		to Fig. \ref{xpx2}, taken at $t=6.48\lambda/c$. $x_{\text{min}}-x_{0}=8,5\cdot10^{-4}\lambda$;
		$x_{\text{max}}-x_{0}=0.05\lambda$; $a=2\cdot10^{-4}$ (simulation
		frame). }
	\label{fit_dens_exp} 
\end{figure}

Of course the generalization $\delta(x)\rightarrow\delta_{a}(x)$
can be also used to calculate the density in the whip case where
$x_{p}(p)=e^{\alpha p}$. But now we have to take $p_{\text{cut}}$ as
a lower limit by the integration (\ref{dens1/x}), since otherwise
the integral would diverge for $x\in[x_{\text{min}}-a,x_{\text{min}}+a]$.
Doing the same steps as in the previous case, we finally obtain the
density profile 
\begin{align}
n_{a}(x)=\left\lbrace \begin{aligned} & \frac{3N}{4a^{3}\ln{\left(\frac{x_{\text{max}}}{x_{\text{min}}}\right)}}\biggr((x+a)\left(x+(x-a)\left(\frac{1}{2}+\ln\left(\frac{x_{\text{min}}}{x+a}\right)\right)\right)\\
 & \qquad\qquad\quad+x_{\text{min}}\left(\frac{1}{2}x_{\text{min}}-2x\right)\biggr)\quad\text{for}\quad x\in[x_{\text{min}}-a,x_{\text{min}}+a]\\
 & \frac{3N}{4a^{3}\ln{\left(\frac{x_{\text{max}}}{x_{\text{min}}}\right)}}\left(2ax-(x^{2}-a^{2})\ln\left(\frac{x+a}{x-a}\right)\right)\quad\text{for}\quad x>x_{\text{min}}+a\\
 & \qquad0\qquad\qquad\qquad\qquad\qquad\qquad\qquad\qquad\qquad\text{for}\quad x<x_{\text{min}}-a
\end{aligned}
\right.\nonumber \\
\label{profile_exp}
\end{align}
Now as in the previous case we are going to compare the calculated
analytical function with the simulated density profile (Fig. \ref{fit_dens_exp}).
Again we obtain good agreement and are able to describe the density
spike quite well.

\begin{figure}[htb]
	\centering \includegraphics[height=7cm]{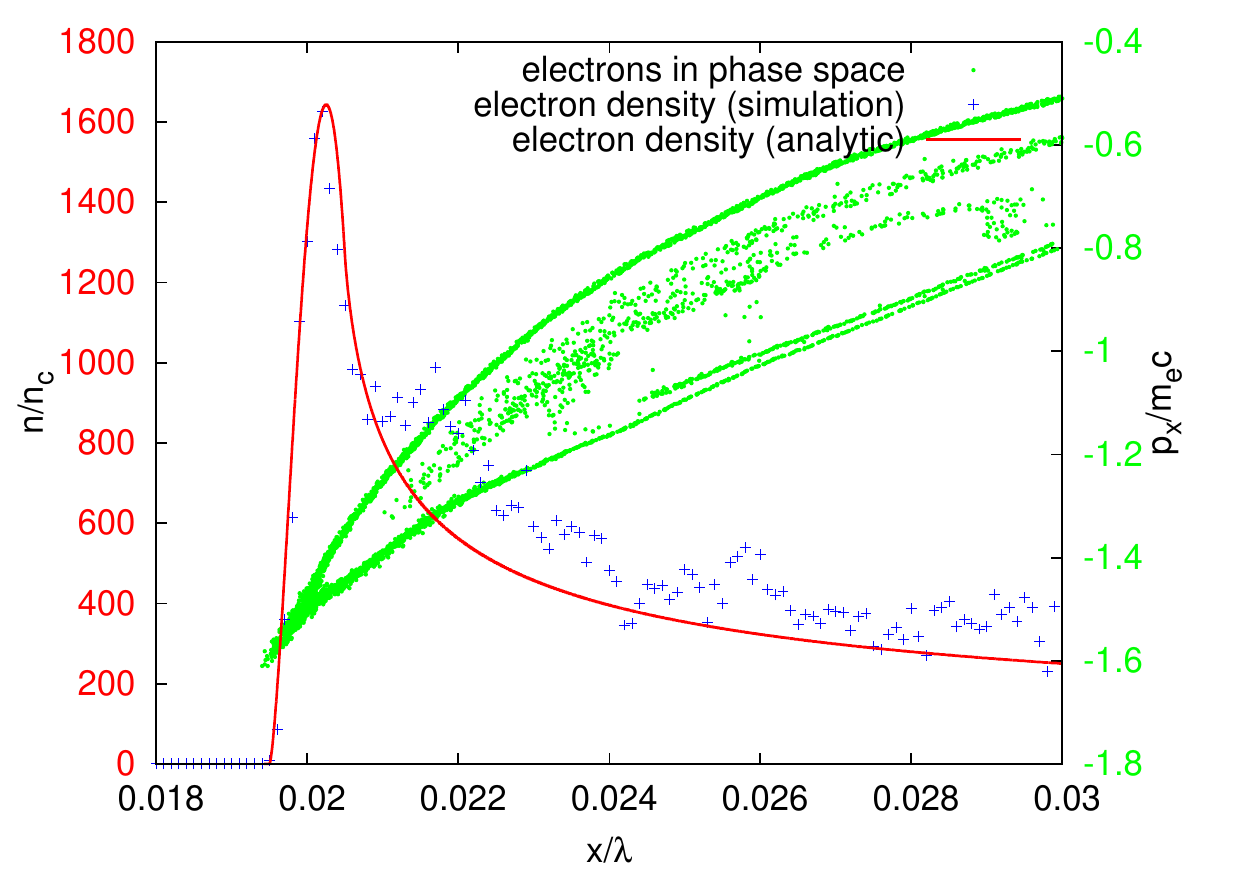} 
	\caption{\small Electron density taken from simulation (blue) and calculated
		analytically via (\ref{profile_sqrt}) (red), as well as electrons
		in $x$-$p_{x}$-plane (green), with same simulation parameters compared
		to Fig. \ref{dens1_377} but to the later time with parameters $\Delta x=0.01\lambda$;
		$a=5\cdot10^{-4}$ (simulation frame).}
	\label{fit_dens_sqrt} 
\end{figure}

Before we go further, we analyze the intermediate case $\dot{x}_{0}(t)\lesssim c$.
In this case, the electron phase space distribution looks like it
is shown in Fig. \ref{fit_dens_sqrt} and can not be approximated well
neither with a parabolic, nor with an exponential function. 
Nevertheless, we find out that the density profile of the spike can
still be well approximated with equation (\ref{profile_sqrt}) (Fig.
\ref{fit_dens_sqrt}), so we classify the cases with intermediate
velocities as parabolic. Fig. \ref{fit_dens_sqrt} is taken from the
same simulation as Fig. \ref{plot_prof1}, but at the later time.

In the following chapter we are going to analyze the corresponding simulation
results more extensively. We will use the descriptions of the electron
layer density profile derived here in order to calculate an expression
for the spectra of the reflected waves in different cases.

\newpage{}

\section{Electron density evolution and HHG emission}
In the previous section, we did the first step towards our aim to improve the analytical description of the spectrum in the case of CSE. 
We derived two analytic expressions which describe the electron density profile in two different cases during CSE process. 
In this section we will go further and work out the equations for the transverse current distribution for the corresponding cases, where the expressions derived previously will be used. 
Since the reflected radiation $E_{r}(t)$ is determined by the transverse current distribution $j_{\bot}(t,x)$ via  
\begin{equation}
E_{r}(t)=\pi\int j_{\bot}(t-x,x)dx,\label{er2}
\end{equation}
where we use the normalized PIC units, see \cite{PIC} for more detail,
the results will lead us directly to the improved description of the spectrum (see below).  
In order to start the derivation, we have to make some assumptions about the current distribution in particular case. 
For that reason we briefly  consider the process of CSE.   

We are interested in the high frequency spectrum of the reflected pulse
mostly determined by the behavior of the electron nanobunch when it
moves away from the plasma with maximal velocity. This moment corresponds
to the stationary phase point (SPP) (see \cite{BP}). The gamma factor
of the bunch exhibits a sharp spike at this time, the so called $\gamma$-spike
\cite{BGP}. One distinguishes different orders of $\gamma$-spikes
depending on behavior of the transverse current (see below). 

\begin{figure}[htb]
	\centering \includegraphics[width=1\textwidth]{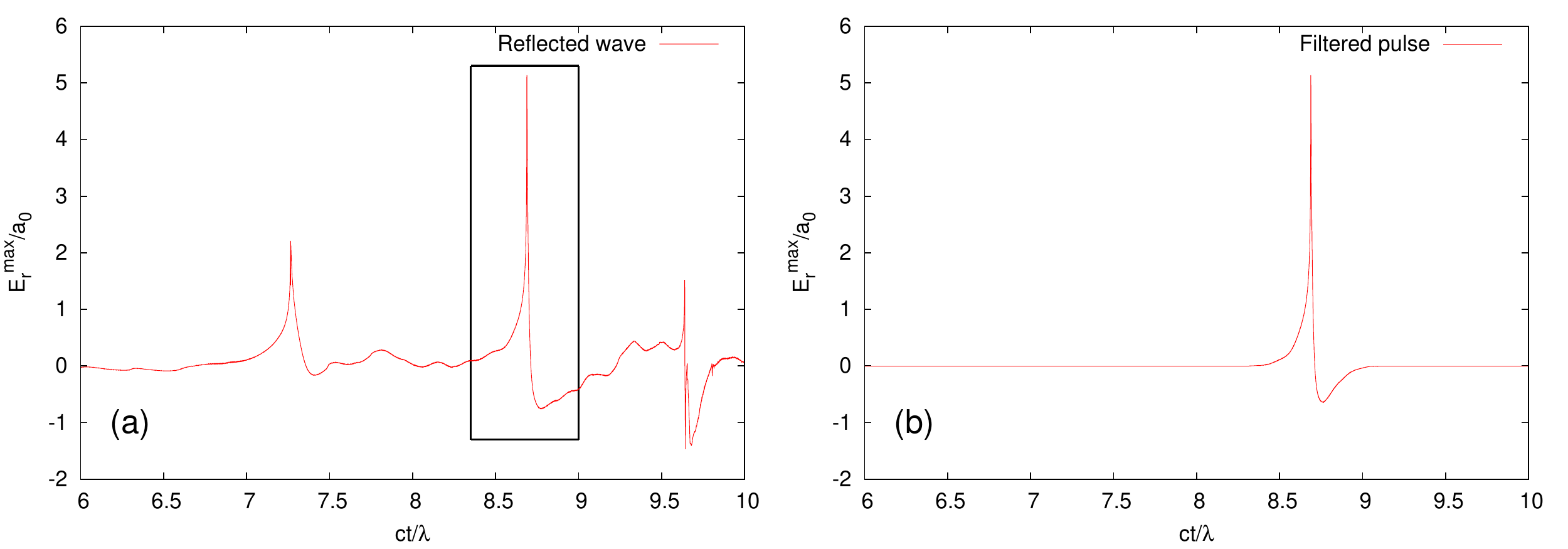} 
	\caption{\small (a): Part of the reflected radiation given by $\frac{1}{2}(E_{y}(t)-B_{z}(t))$.
		(b): Single pulse from the reflected pulse train filtered out by the
		Gaussian function. Simulation parameters are the same as for
		Fig. \ref{xpx2}.}
	\label{er} 
\end{figure}

\begin{figure}[htb]
	\centering \includegraphics[width=1\textwidth]{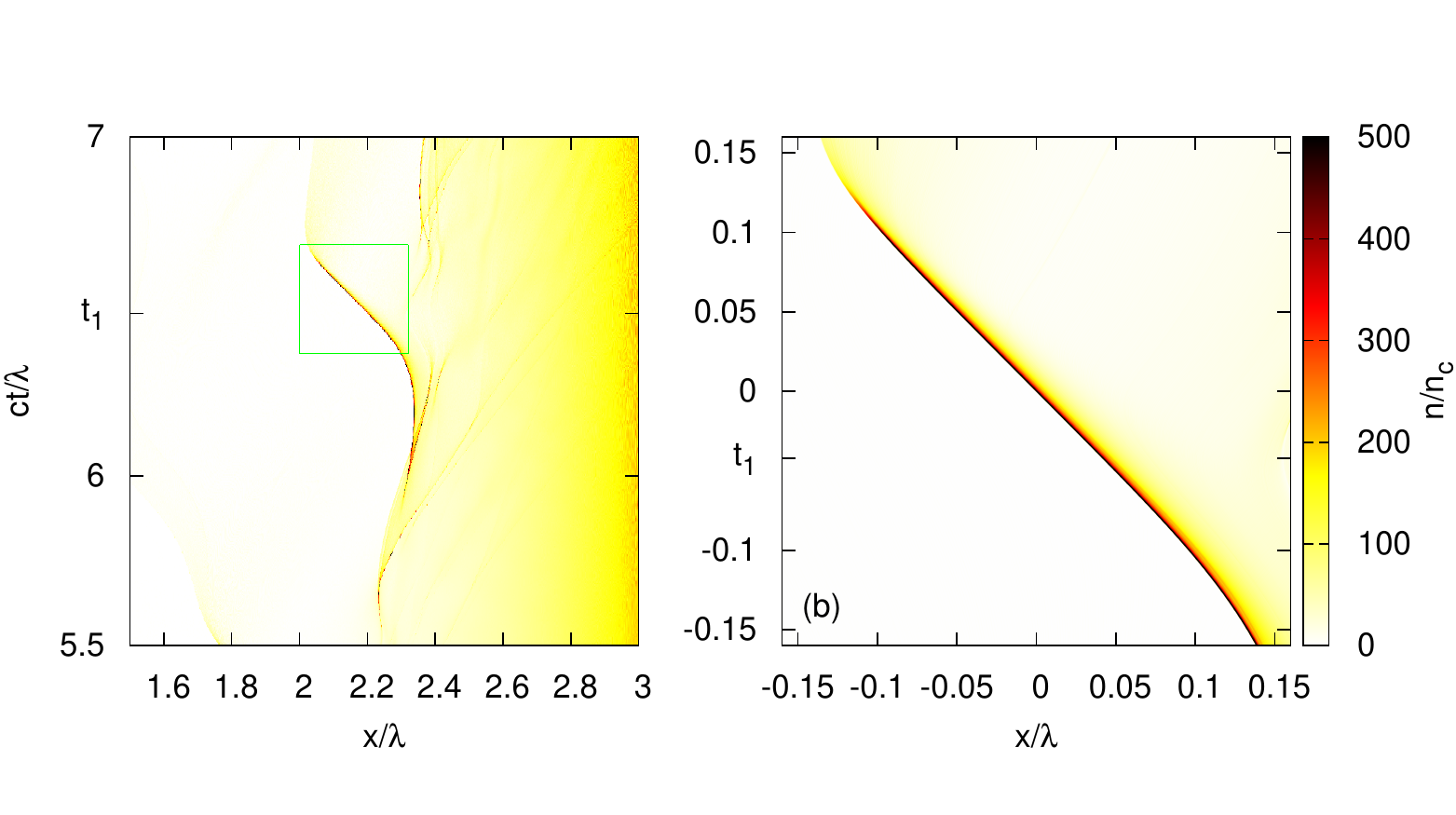} 
	\caption{\small The electron density distribution in space time domain.
		Simulation parameters are the same as for Fig. \ref{xpx2}. All
		magnitudes are taken in the simulation frame The green square in (a)
		marks the nanobunch that is zoomed in (b). This bunch is responsible
		for the radiation of the strong pulse shown in Fig. \ref{er}. In
		(a) $t_{1}=6.48\lambda/c$ as in Fig. \ref{fit_dens_exp}.}
	\label{dens1_2D} 
\end{figure}

First, we investigate the example of the whip case ($\dot{x}_{0}(t)\rightarrow c$)
from the previous section illustrated in Fig. \ref{dens_comp}(b),
Fig. \ref{xpx2} and Fig. \ref{fit_dens_exp} more extensively. The
reflected wave obtained in this simulation is shown in Fig. \ref{er}
(a). Since we use a few cycle laser pulse, we get an attosecond pulse
train as reflected radiation. In the following, we consider only the
most intense reflected pulse $E_{r}^{\text{pls}}(t)$ that is filtered
out by the Gaussian function (Fig. \ref{er} (b)), i.e. 
\begin{equation}
E_{r}^{\text{pls}}(t)=E_{r}(t)e^{(t-t_{\text{max}})^{2}/\tilde{\sigma}^{2}},
\end{equation}
where $t_{\text{max}}$ corresponds to the maximal wave amplitude
and $\tilde{\sigma}=0.2\lambda/c$. The amplitude of the pulse is
about five times larger than of incident wave like in the CSE case.
The electron nanobunch which radiates this pulse can be clearly recognized
from the density distribution shown in Fig. \ref{dens1_2D}. 
For convenience, we chose the coordinates in Fig. \ref{dens1_2D}
(b) in such a way that the SPP is in the point (0,0), while in Fig.
\ref{dens1_2D} (a) it corresponds to the point ($2.16\lambda$, $6.52\lambda/c$).

\begin{figure}[t]
	\centering \includegraphics[width=1\textwidth]{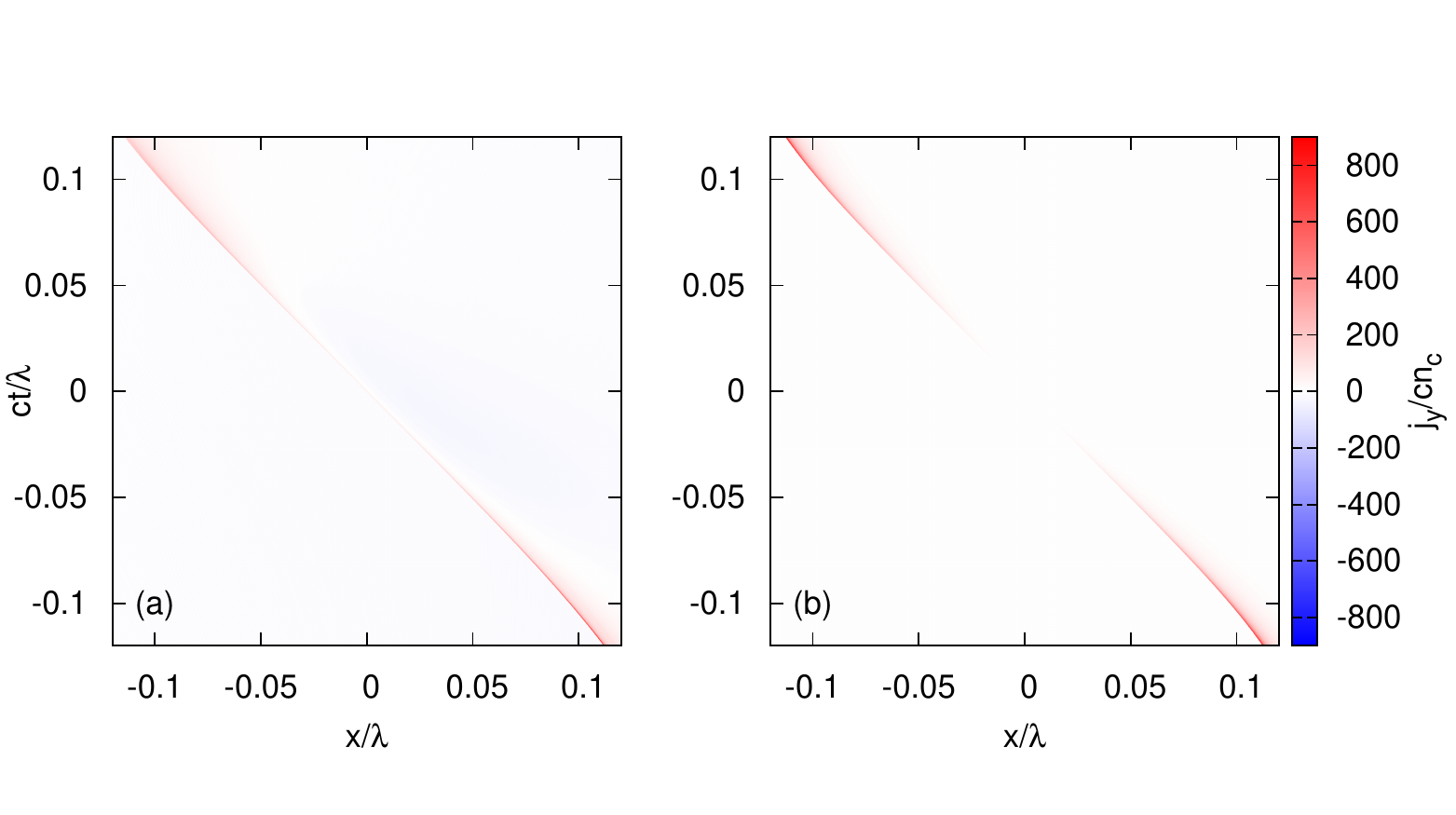} 
	\centering \includegraphics[width=1\textwidth]{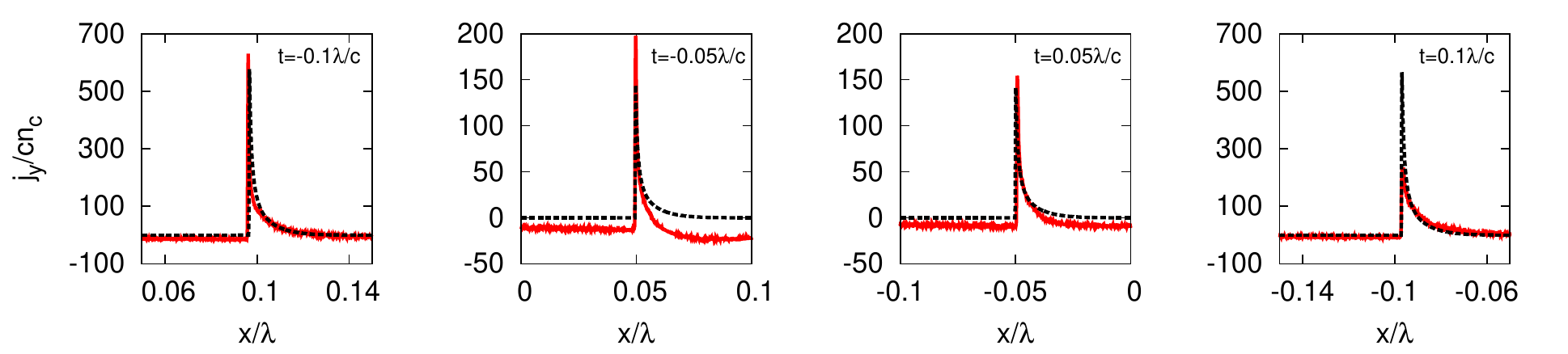} 
	\caption{\small Transverse current density from the simulation (a) and
		calculated analytically (b). In (a) the simulated current density
		near the SPP (0,0) is illustrated. As we see the transverse current
		almost vanish in SPP. Simulation parameters are the same compared
		to Fig. \ref{xpx2}. In (b) the analytically calculated current distribution
		is shown. The parameters used in equation (\ref{x_0}) are: $\alpha_{0}=6\cdot10^{4}$,
		$n_{\text{m}}=1100$ and $\gamma=15$, while the parameters used for
		the shape are: $a=1,5\cdot10^{-4}\lambda$, $x_{\text{min}}=8\cdot10^{-4}\lambda$
		and $\tilde{\tilde{\sigma}}=0.02\lambda$. The velocity $\upsilon$
		in (\ref{x_0}) is derived from the given gamma factor. 
		Bottom pictures illustrate the simulated (red) and calculated (black) current at the reference times $-0.1\lambda/c$, $-0.05\lambda/c$, $0.05\lambda/c$ and $0.1\lambda/c$.}
	\label{jy_2D} 
\end{figure}

Now we are able to make some assumption of current behavior in the vicinity of the SPP. 
Our derivation is similar to the one in \cite{BP}, but is more detailed.
First of all, we assume that the transverse current density does not
change its shape during the time and write 
\begin{equation}
j_{\bot}(t,x)=n(t,x)v_{\bot}(t,x)\approx j(t)f(x-x_{0}(t)),\label{j1}
\end{equation}
with density $n(t,x)$, velocity $v_{\bot}(t,x)$ and the position
of the bunch $x_{0}$. The function $f$ is the shape that is assumed
to be constant close to the SPP. If we compare the both sides of the
last equation in (\ref{j1}), we can approximate  
\begin{align}
\upsilon_{\bot}(t,x) & \approx\bar{\upsilon}_{\bot}(t),\qquad n(t,x)\approx n_{\text{m}}f(x-x_{0}(t)).\label{assumptions}
\end{align}
This means that equation (\ref{j1}) assumes the transverse velocity
being approximately constant in space and equal to the mean velocity
$\bar{\upsilon}_{\bot}(t)$. The density profile has a constant shape
with the maximal value $n_{\text{m}}$ and only changes its position
$x_{0}(t)$. Thus the equation (\ref{j1}) takes the form 
\begin{align}
j_{\bot}(t,x) & \approx n_{\text{m}}\bar{v}_{\bot}(t)f(x-x_{0}(t)),\qquad j(t)=n_{\text{m}}\bar{v}_{\bot}(t).\label{j}
\end{align}
In the next step, we assume the ultrarelativistic regime, which means
that absolute velocity of the particles is always close to $c$. In
this case we can write 
\begin{equation}
\sqrt{\dot{x}_{0}(t)^{2}+\bar{v}_{\bot}(t)^{2}}\approx\upsilon\approx c\label{relat}
\end{equation}
Now we are going to Taylor expand $j(t)$. Let us first take a look
at the Fig. \ref{jy_2D} (a). 
It shows the transverse current distribution of the given nanobunch.
In the SPP the bunch exhibits maximal longitudinal velocity $\upsilon_{x}=\upsilon$
which is close to $c$, so that the transverse component almost vanish.
As a result, the transverse current vanishes as well. This can be seen
in the picture. We can also see, that the current does not change its
sign at the SPP. So we assume 
\begin{equation}
j(t)\approx\alpha_{0}t^{2}.\label{j_teylor}
\end{equation}
Combining this relation with the second equation from (\ref{j}) and
inserting it in (\ref{relat}) leads to 
\begin{align}
\dot{x}_{0}(t) & \approx-\sqrt{\upsilon^{2}-\frac{\alpha_{0}^{2}}{n_{\text{m}}^{2}}t^{4}}\approx-\upsilon+\frac{\alpha_{0}^{2}}{2\upsilon n_{\text{m}}^{2}}t^{4},\nonumber \\
x_{0}(t) & \approx-\upsilon t+\frac{\alpha_{0}^{2}}{2\upsilon n_{\text{m}}^{2}}\frac{t^{5}}{5}\equiv-\upsilon t+\alpha_{1}\frac{t^{5}}{5}.\label{x_0}
\end{align}
Here we have the negative sign in front of the square root since the
electron layer moves in the negative direction. To describe the shape
$f$ we use the expression for the density profile (\ref{profile_exp})
which is derived in the previous section. With this function we replace
the Gaussian function that is used in \cite{BP}. Obviously the function
(\ref{profile_exp}) fits the density much better than the Gaussian
function as shown above. From the second equation of (\ref{assumptions}),
we read that the maximal value of $f$ is one. Consequently we conclude, that
\[
f(x)=\frac{1}{n_{\text{m}}}n_{a}(x).
\]
Since $n_{\text{m}}$ should represent the maximum of $n_{a}$, we
can write $n_{\text{m}}=n_{a}(x_{\text{m}})$, where $x_{\text{m}}$
is the extrem value of $n_{a}$, which depends on parameters $a$
and $x_{\text{min}}$. In addition we multiply the shape function
with a wider Gaussian function since $f$ decays too slowly ($\propto1/x$)
for positive $x$ and after certain $x$-value does not coincide with the
given density. Thus the Gaussian helps as to cut this ``tail'' with
no influence on the spectrum structure. So we have 
\begin{equation}
f(x)=\frac{n_{a}(x)}{n_{a}(x_{\text{m}})}e^{-\frac{x^{2}}{\tilde{\tilde{\sigma}}^{2}}}.\label{f}
\end{equation}
Now, we just need to insert the equations (\ref{j_teylor}), (\ref{x_0})
and (\ref{f}) in (\ref{j1}) in order obtain the analytical expression
of the current distribution. To get some result we need to choose
the parameters contained in this formulas, in the way that the calculated
distribution would be similar to those we obtained from simulation
(Fig. \ref{jy_2D} (a)). Moreover, the physical values like the maximal
density $n_{\text{m}}$ or maximal velocity $\upsilon$ have to be
in line with the simulation results. We find that the parameter set
we used to obtain Fig. \ref{jy_2D} (b) is a good choise. 
\begin{figure}[htb]
	\centering 
	\includegraphics[height=7cm]{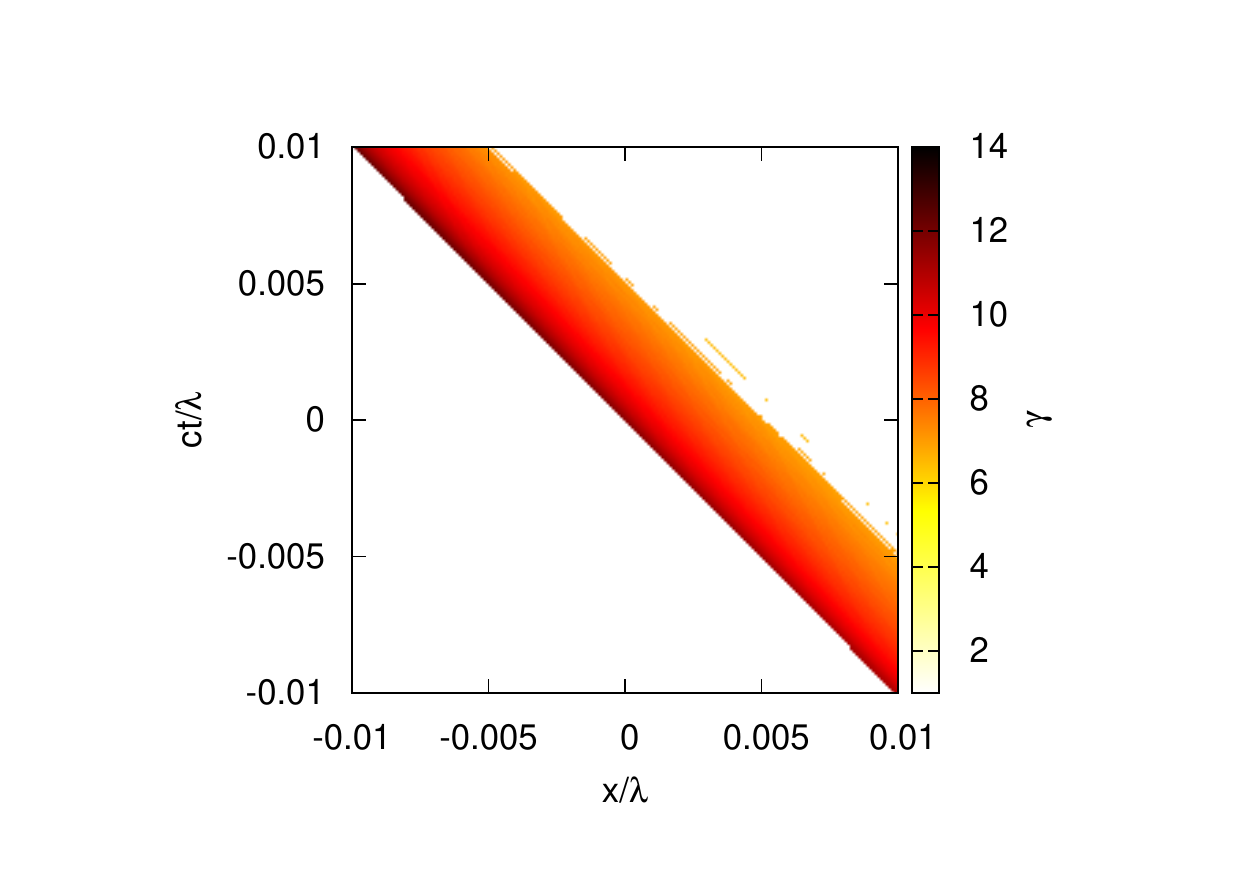} \caption{\small Distribution of $\gamma$ factor in the given nanobunch
		 calculated only in cells with density bigger than 200$n_{c}$.}
	\label{gamma_2D} 
\end{figure}

First of all, we see that the analytical current distribution fits
the original one quite well (Fig. \ref{jy_2D}). 
From the bottom pictures in Fig. \ref{jy_2D} we see that the width and the trajectory of the current peak are well predicted by the analytical model.                   
The corresponding maximums can considerably deviate at some points like at $t=-0.05\lambda/c$ and $t=0.1\lambda/c$ but still remain at the same order of magnitude.     
The current also does not vanish completely at the SPP as expected for the ideal case, but reaches the minimum of about $60$cn$_c$.                                                 
The chosen maximal density $n_{\text{m}}$ is in line with simulation as can be seen
from Fig. \ref{fit_dens_exp}. In order to have an idea concerning
the order of magnitude of gamma factor, we visualized the distribution
of $\gamma$ in the vicinity of density spike (Fig. \ref{gamma_2D}).
As we see $\gamma$ almost reaches the value $\gamma=15$. The numbers
$a$ and $x_{\text{min}}$ that characterize the shape are similar
to those we used in Fig. \ref{fit_dens_exp}. They are slightly different
because these numbers pass better for fitting the current density
through some finite time interval, while in case of Fig. \ref{fit_dens_exp}
only one particular time point is considered.

\begin{figure}[htb]
	\centering 
	\includegraphics[width=1\textwidth]{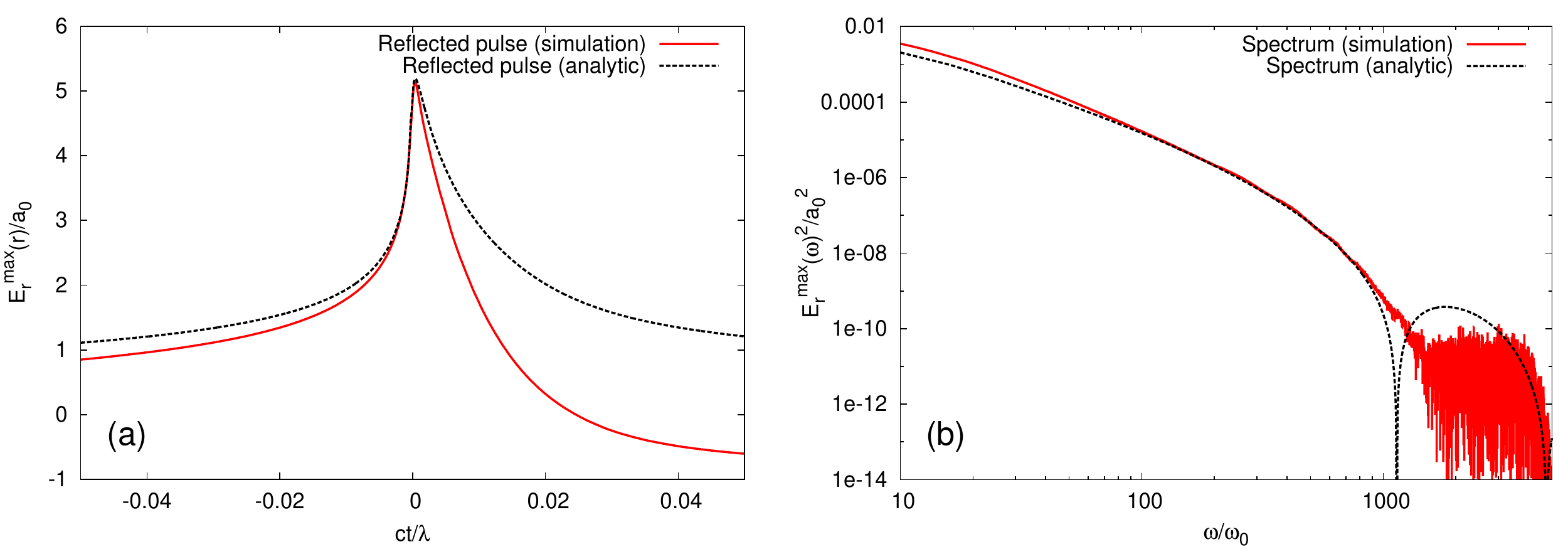} 
	\caption{\small Reflected radiation obtained from the simulation ((a)
		red) and from analytical current distribution ((a) black), as well
		as the corresponding spectra in (b). The spectrum from the simulation
		is taken directly from the radiated pulse via FFT, while the other
		one is obtained using the equation (\ref{Ai2}). }
	\label{airy} 
\end{figure}

Now we consider the radiation emission from the assumed current distribution
as well as its spectrum. Equation (\ref{er2}) enables us to calculate
the radiation and the spectrum analytically. Here we give the expression
of radiated spectrum that is derived in line with \cite{BP}. 
\begin{align}
I(\omega)=E_{r}^{2}(\omega)=4\pi^{4}\alpha_{0}^{2}(\alpha_{1}\omega)^{-\frac{6}{5}}\left(Ai_{2}''(\alpha_{1}^{-\frac{1}{5}}\delta\omega^{\frac{4}{5}})\right)^{2}|f(\omega)|^{2},\label{Ai2}\\
\alpha_{1}=\frac{a_{0}^{2}}{2\upsilon n^{2}},\quad\delta=1-\upsilon,\quad Ai_{2}''=\frac{d^{2}}{dx^{2}}\frac{1}{2\pi}\int e^{i\left(xt+\frac{t^{5}}{5}\right)}dt.\nonumber 
\end{align}
The Fourier transform of the shape function $f(\omega)$ is calculated
numerically using FFT. In Fig. \ref{airy} (b) the spectrum calculated
using (\ref{Ai2}) is compared with the spectrum calculated from original
reflected pulse (Fig. \ref{er}) via FFT. Obviously the description
works very well almost until 1000-th harmonic. The both graphs diverge
for $\omega<100\omega_{0}$ but anyway the method of SPP used here
works only for high harmonics, so we may not expect the coincidence
for low frequencies. 
In Fig. \ref{airy} (a) the corresponding pulses are compared. So
the red one is the same as shown in Fig. \ref{er} and the black one
is determened from the assumed analytical current distribution using
(\ref{er2}). The bough graphs behave in the similar manner. To conclude,
we can say that we obtained quite good results applying our new shape
function derived in the previous section instead of a Gaussian function.

\begin{figure}[htb]
	\centering 
	\includegraphics[width=1\textwidth]{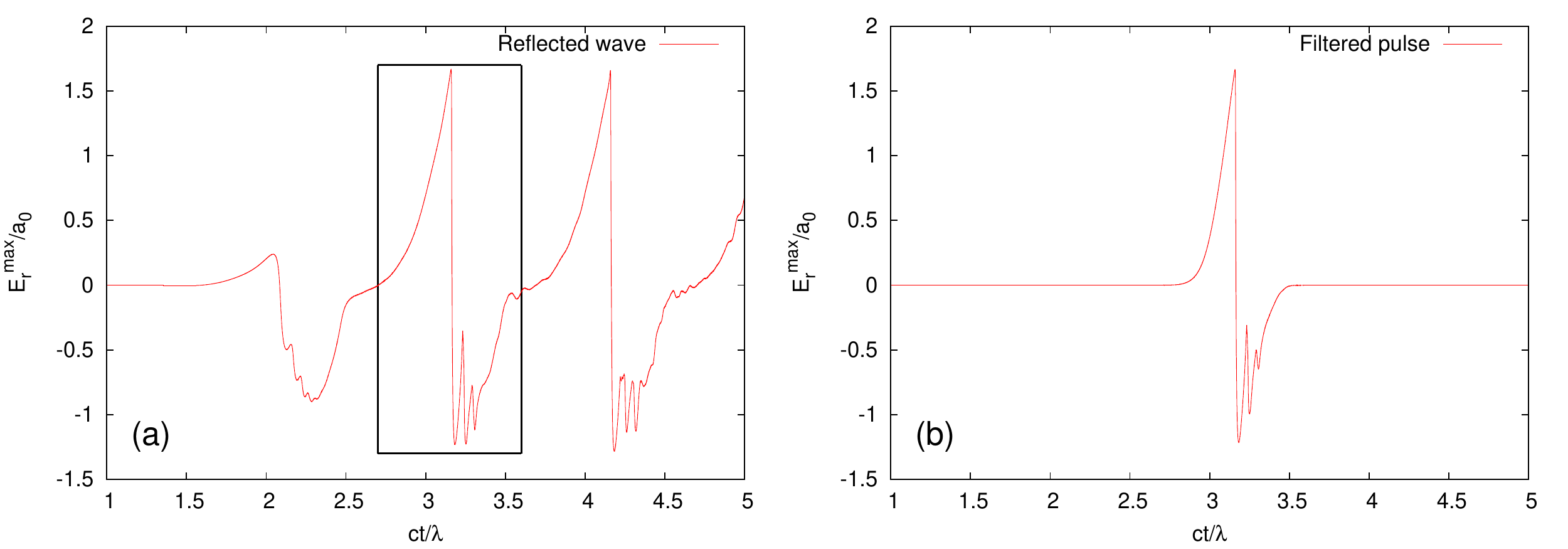} 
	\caption{\small (a) Part of the reflected radiation given by $\frac{1}{2}(E_{y}(t)-B_{z}(t))$.
		(b) Single pulse from the reflected pulse train filtered out by the
		Gaussian function. Simulation parameters are the same as for
		Fig. \ref{dens1_377}.}
	\label{er2.2} 
\end{figure}

\begin{figure}[htb]
	\centering 
	\includegraphics[width=1\textwidth]{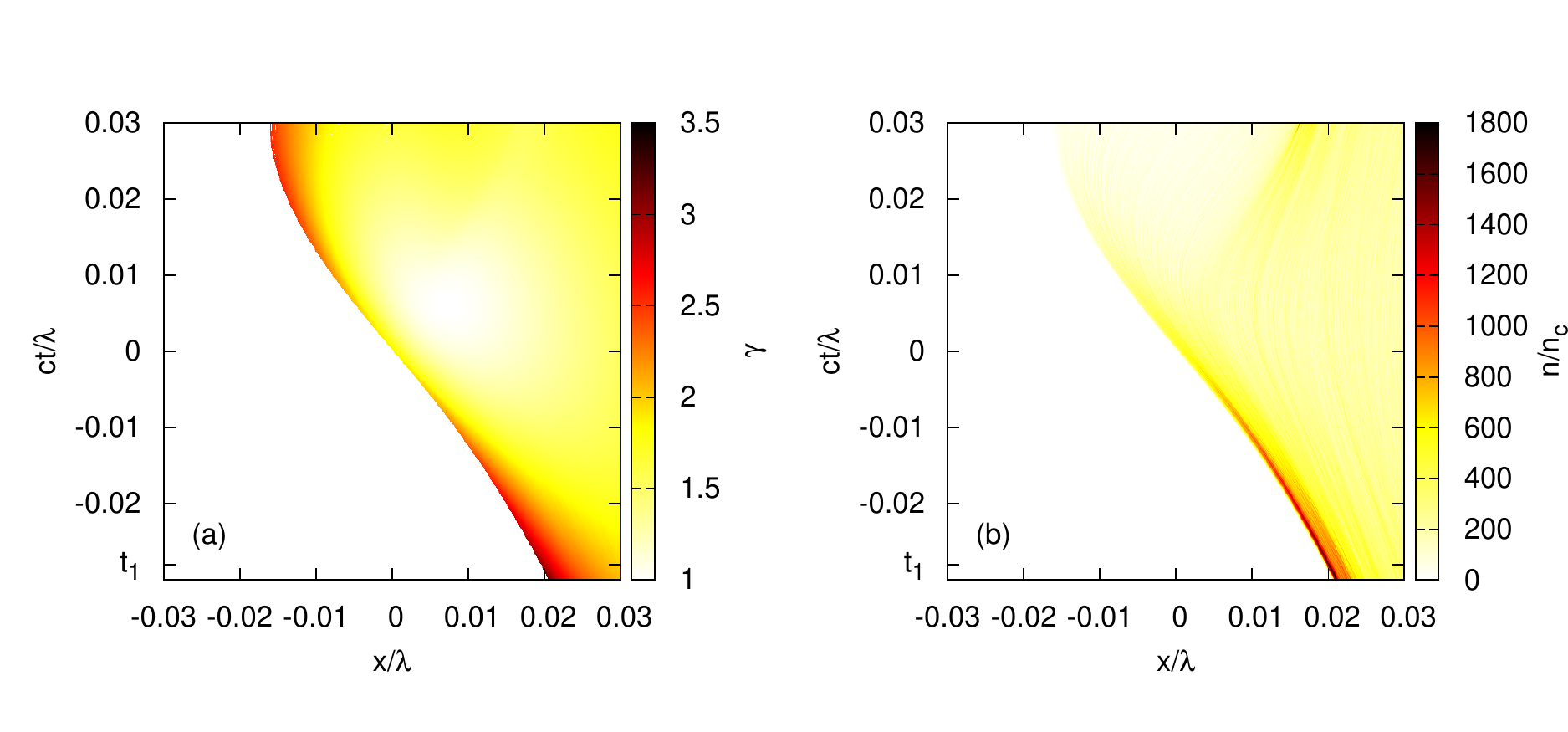} 
	\caption{\small (a) Distribution of the gamma factor in the given electron
		bunch in space time domain. (b) The electron density distribution
		in space time domain. Simulation parameters are the same compared
		to Fig. \ref{dens1_377}. All magnitudes are taken in the simulation
		frame. $t_{1}$ denotes the time which corresponds to Fig. \ref{fit_dens_sqrt},
		$t_{1}=-0.028\lambda/c$. }
	\label{dens1_gamma_2D} 
\end{figure}

Going along the same line we analyze now the intermediate case $\dot{x}_{0}(t)\lesssim c$
shown in Fig. \ref{fit_dens_sqrt}. As said before, we attribute
this case to the parabolic case. First of all, we consider the reflected
pulse train and filter out the pulse we are interested in (Fig. \ref{er2.2}).
The density and the gamma factor of the corresponding electron bunch
is shown in Fig. \ref{dens1_gamma_2D}. 
The point (0,0) corresponds to the SPP like in the previous case. Even
from this picture one can clearly see that the velocity in the SPP significantly
deviates from the speed of light and is approximately $0.85c$.
From the distribution of the gamma factor we see that it is roughly
constant within the electron layer. So we can use the same approximation
as in the previous case 
\begin{equation}
\sqrt{\dot{x}_{0}(t)^{2}+\bar{v}_{\bot}(t)^{2}}\approx\upsilon.\label{relat2}
\end{equation}
The difference between (\ref{relat}) and (\ref{relat2}) is of course
that in the last equation $\upsilon$ is not close to $c$. For that
reason the electron phase space distribution does not become ``whip-like''
(Fig. \ref{fit_dens_sqrt}). 
\begin{figure}[htb]
	\centering \includegraphics[width=1\textwidth]{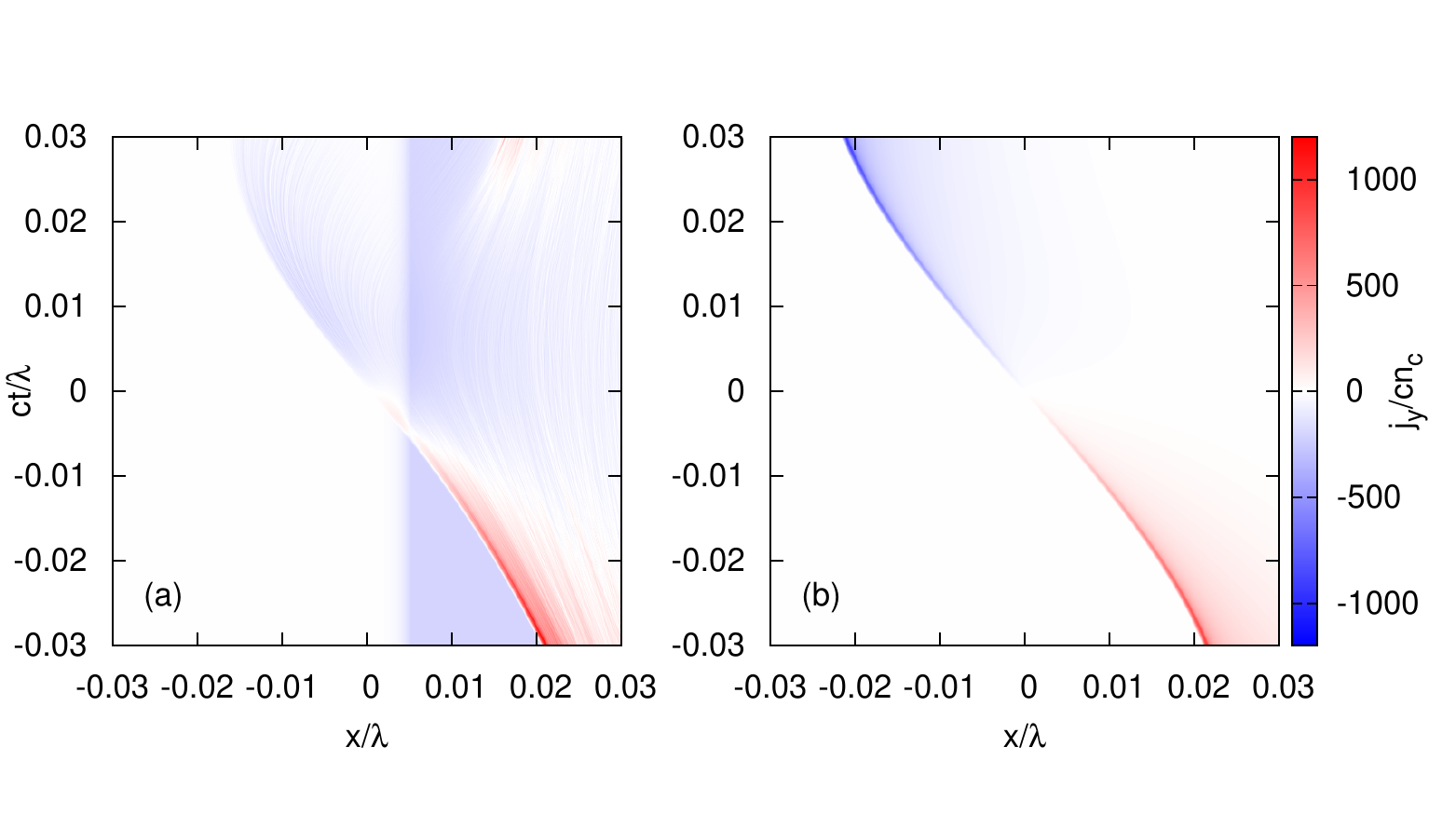} 
	\centering \includegraphics[width=1\textwidth]{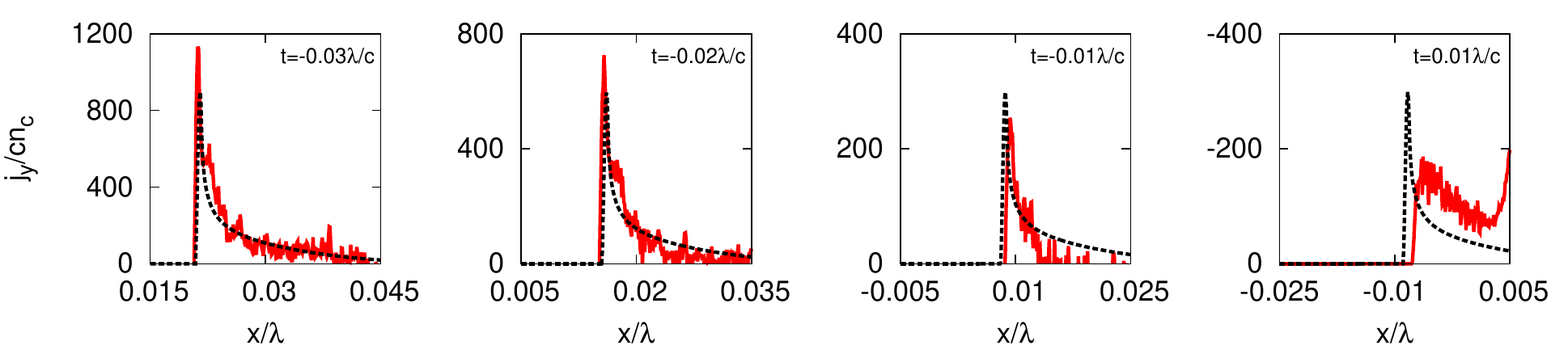}
	\caption{\small Transverse current density from the simulation (a) and
		calculated analytically (b). In (a) the simulated current density
		near the SPP is illustrated. Simulation parameters are the same compared
		to Fig. \ref{dens1_377}. In (b) the analytically calculated current
		distribution is shown. The parameters used in equation (\ref{x_02})
		are: $\alpha_{0}=3\cdot10^{4}$, $n_{\text{m}}=1000$ and $\gamma=2$,
		while the parameters used for the shape are: $a=4\cdot10^{-4}\lambda$
		and $\tilde{\tilde{\sigma}}=0.02\lambda$. The velocity $\upsilon$
		in (\ref{x_0}) is derived from the given gamma factor.
		Bottom pictures illustrate the simulated (red) and calculated (black) current at the reference times $-0.03\lambda/c$, $-0.02\lambda/c$, $-0.01\lambda/c$ and $0.01\lambda/c$.}
	\label{jy2_2D}
\end{figure}

\begin{figure}[ht]
	\centering 
	\includegraphics[width=1\textwidth]{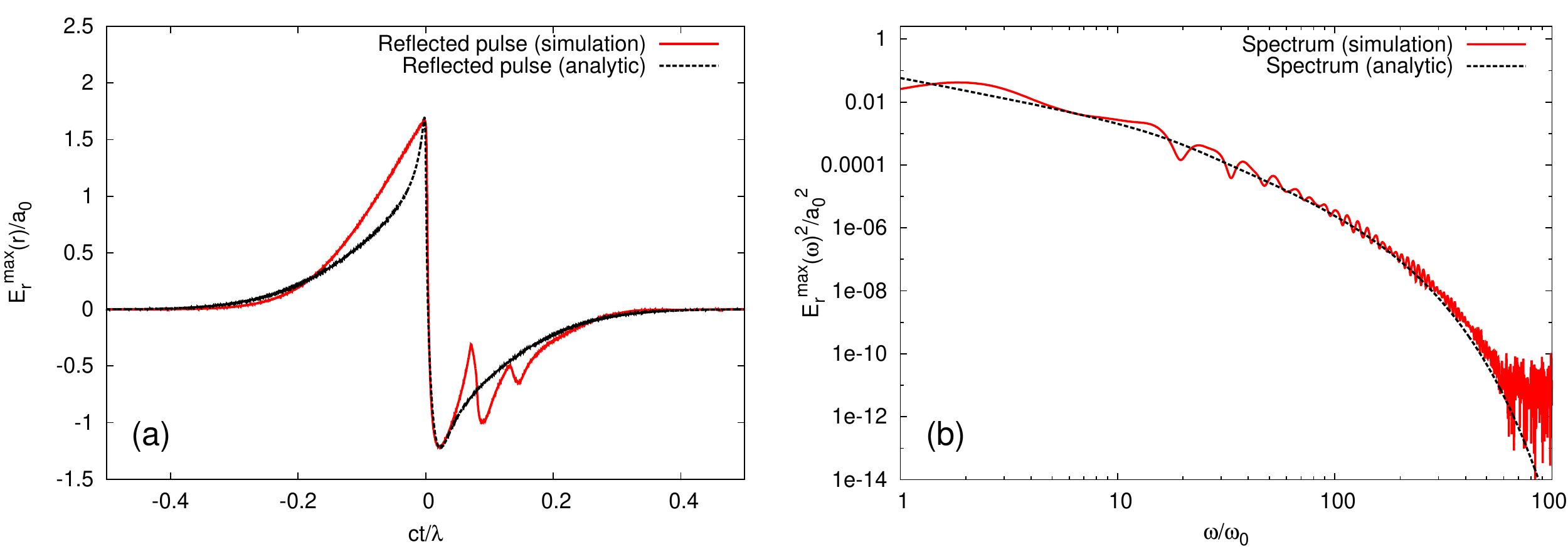} 
	\caption{\small Reflected radiation obtained from the simulation ((a)
		red) and from analytical current distribution ((a) black), as well
		as the corresponding spectra in (b). The spectrum from the simulation
		is taken directly from the radiated pulse via FFT, while the other
		one is obtained using the equation (\ref{Ai1}). }
	\label{airy1} 
\end{figure}

Now let us consider the corresponding current distribution (Fig. \ref{jy2_2D}(a)).
In this case, the current changes its sign in the SPP, so we can assume
\begin{equation}
j(t)\approx-\alpha_{0}t.\label{j_teylor2}
\end{equation}
Similar to the previous case, we derive 
\begin{equation}
x_{0}(t)\approx-\upsilon t+\frac{\alpha_{0}^{2}}{2\upsilon n_{\text{m}}^{2}}\frac{t^{3}}{3}\equiv-\upsilon t+\alpha_{1}\frac{t^{3}}{3}.\label{x_02}
\end{equation}
Using these assumptions we calculate the current distribution analytically
(Fig. \ref{jy2_2D}(b)).  
In this case we obtain a good agreement only for negative times (see bottom pictures of Fig. \ref{jy2_2D}), but the predicted trajectory of the peak is still close to original one.  
The gamma factor can be roughly read from
Fig. \ref{dens1_gamma_2D}, so we set $\gamma=2$. This value matches
well the maximum longitudinal velocity of the layer obtained above.
Finally we obtain the spectrum of the reflected wave 
\begin{align}
I(\omega)=E_{r}^{2}(\omega)=4\pi^{4}\alpha_{0}^{2}(\alpha_{1}\omega)^{-\frac{4}{3}}\left(Ai_{1}'(\alpha_{1}^{-\frac{1}{3}}\delta\omega^{\frac{2}{3}})\right)^{2}|f(\omega)|^{2},\label{Ai1}\\
\alpha_{1}=\frac{a_{0}^{2}}{2\upsilon n^{2}},\quad\delta=1-\upsilon,\quad Ai_{1}'=\frac{d}{dx}\frac{1}{2\pi}\int e^{i\left(xt+\frac{t^{3}}{3}\right)}dt.\nonumber 
\end{align}
The corresponding pulses and their spectra are shown in Fig. \ref{airy1}.
Even if the above assumptions did not work as good as in the previous
case, the obtained results still give a satisfactory approximation.

We can generalize the equations (\ref{Ai2}) and (\ref{Ai1}) and
write: 
\begin{align*}
I(\omega) & =E_{r}^{2}(\omega)=4\pi^{4}\alpha_{0}^{2}(\alpha_{1}\omega)^{-\frac{2n+2}{2n+1}}\left(\frac{d^{n}}{d\xi^{n}}Ai_{n}(\xi_{n})\right)^{2}|f(\omega)|^{2},\\
\xi_{n} & =\alpha_{1}^{-\frac{1}{2n+1}}\delta\omega^{\frac{2n}{2n+1}},\quad Ai_{n}=\frac{1}{2\pi}\int e^{i\left(xt+\frac{t^{2n+1}}{2n+1}\right)}dt.
\end{align*}
We obtain this formula from the general assumption $j(t)=\alpha_{0}(-t)^{n}$
for the transverse current. The index $n$ corresponds to the order
of a certain $\gamma$-spike. Thus, we see that in the first example
(whip case) we have the second order gamma spike, while in the second
example (parabolic case) the first order gamma spike is obtained.

\begin{figure}[htb]
	\centering 
	\includegraphics[width=1\textwidth]{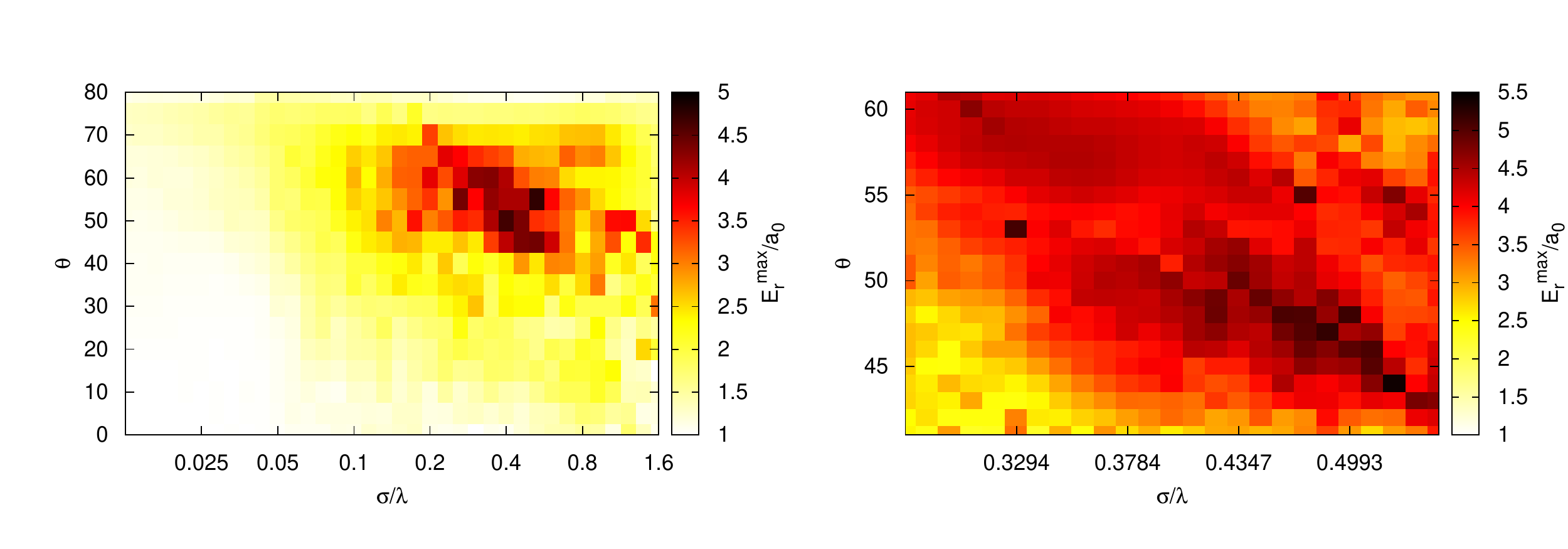} 
	\caption{\small Each point in these pictures corresponds to the maximal
		amplitude obtained from the reflected radiation taken from corresponding
		simulation. We have different angles of incidence along $y$-axis
		and different steepness of the density gradient along $x$-axis, where
		$\sigma$ is taken from (\ref{rump}) and $n_{0}=100n_{c}$ (laboratory
		frame). The right picture is the zoom in the parameter range where
		the most strong amplification is obtained. }
	\label{params} 
\end{figure}

Now it is time to deal with our second goal, namely  to investigate the most efficient case of HHG at moderate laser intensity ($a_{0}=10$). 
For this purpose we perform several 1D PIC simulations and vary the steepness of
the exponential density gradient as well as incident angle. For each
parameter set we consider the reflected radiation in order to find
the increase of the amplitude that is common in the case of nanobunching.
In Fig. \ref{params} we visualized the maximal amplitude of the reflected
wave for each parameter set respectively. 
Consider the incident angle between 45 and 60, since by this angles
the most interesting things happen. Of course, one notices the sharp
increase of the reflected wave amplitude in the area around $\sigma=0.4\lambda$
zoomed in Fig. \ref{params}(b). By $\sigma=0.5\lambda$ and the
angle $48^{\circ}$ (laboratory frame) we get the amplification of a
factor of five. This is the most efficient HHG we could obtain. We call this area high amplitude parameter set (HAPS).
In this area we mostly obtain the second order $\gamma$-spikes and
the current does not change its sign in the SPPs like in Fig. \ref{jy_2D}.
Furthermore, our study shows that the maximum longitudinal velocity
of the boundary electron layer increases monotonically with
$\sigma$ until HAPS, where it almost reaches $c$. For $\sigma<0.05\lambda$
the boundary oscillates too slowly so that no short pulses are generated.
Roughly in the range between $0.05\lambda$ and $0.1\lambda$ we obtain
the reflected radiation very similar to that from Fig. \ref{er2.2}
and generated via the same mechanism. We call this area moderate amplitude
parameter set (MAPS). Here we have only first order $\gamma$-spikes
and the current changes sign in the SPPs. Thus the reflected spectrum
in MAPS can be approximated with equation (\ref{profile_sqrt}) (parabolic
case) and the area of HAPS corresponds to the exponential case
(equation (\ref{profile_exp})). In the area between MAPS and HAPS
the interaction is too complicated to be attributed to any model.

\section{Transmitted radiation in nanobunching regime}

\begin{figure}[htb]
	\centering 
	\includegraphics[width=1\textwidth]{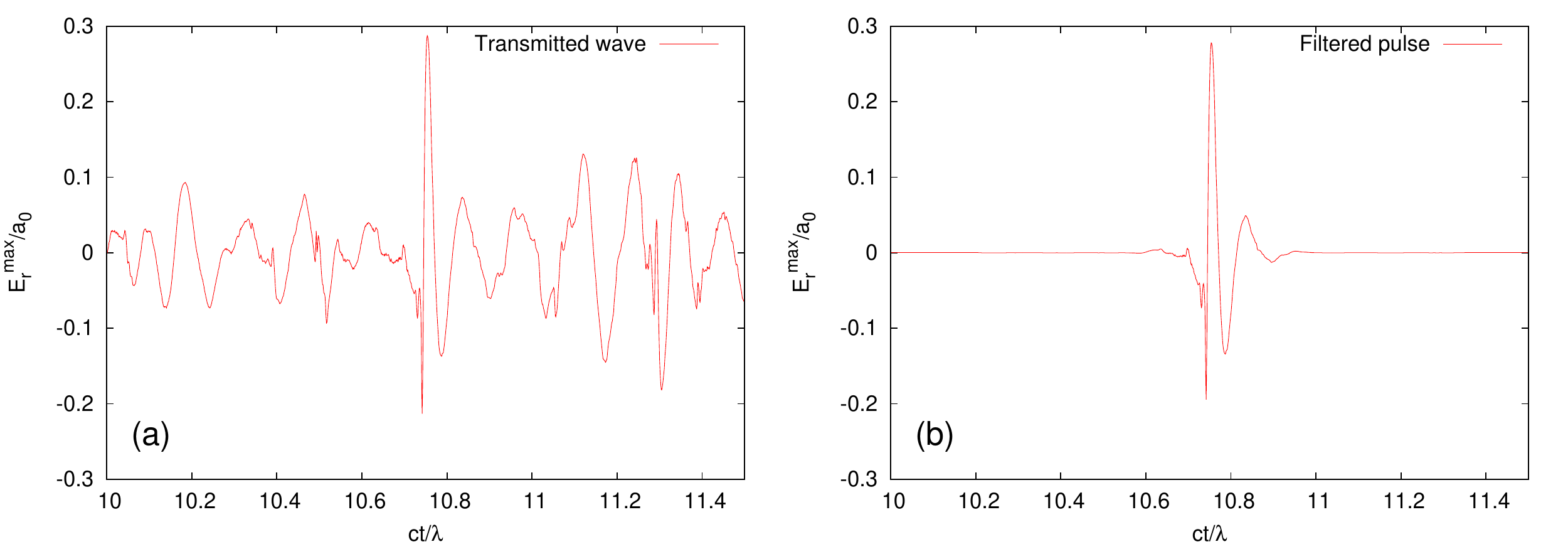} 
	\caption{\small (a) Part of the transmitted radiation. (b) Single pulse
		filtered out by the Gaussian function. Simulation parameters: initial
		plasma density $n_{0}=40n_{c}$; $\sigma=0.4\lambda$, normal incident
		pulse with dimensionless amplitude $a_{0}=10$ has the wave length
		$\lambda=820$nm.}
	\label{ei}  
\end{figure}

\begin{figure}[htb]
	\centering 
	\includegraphics[width=1\textwidth]{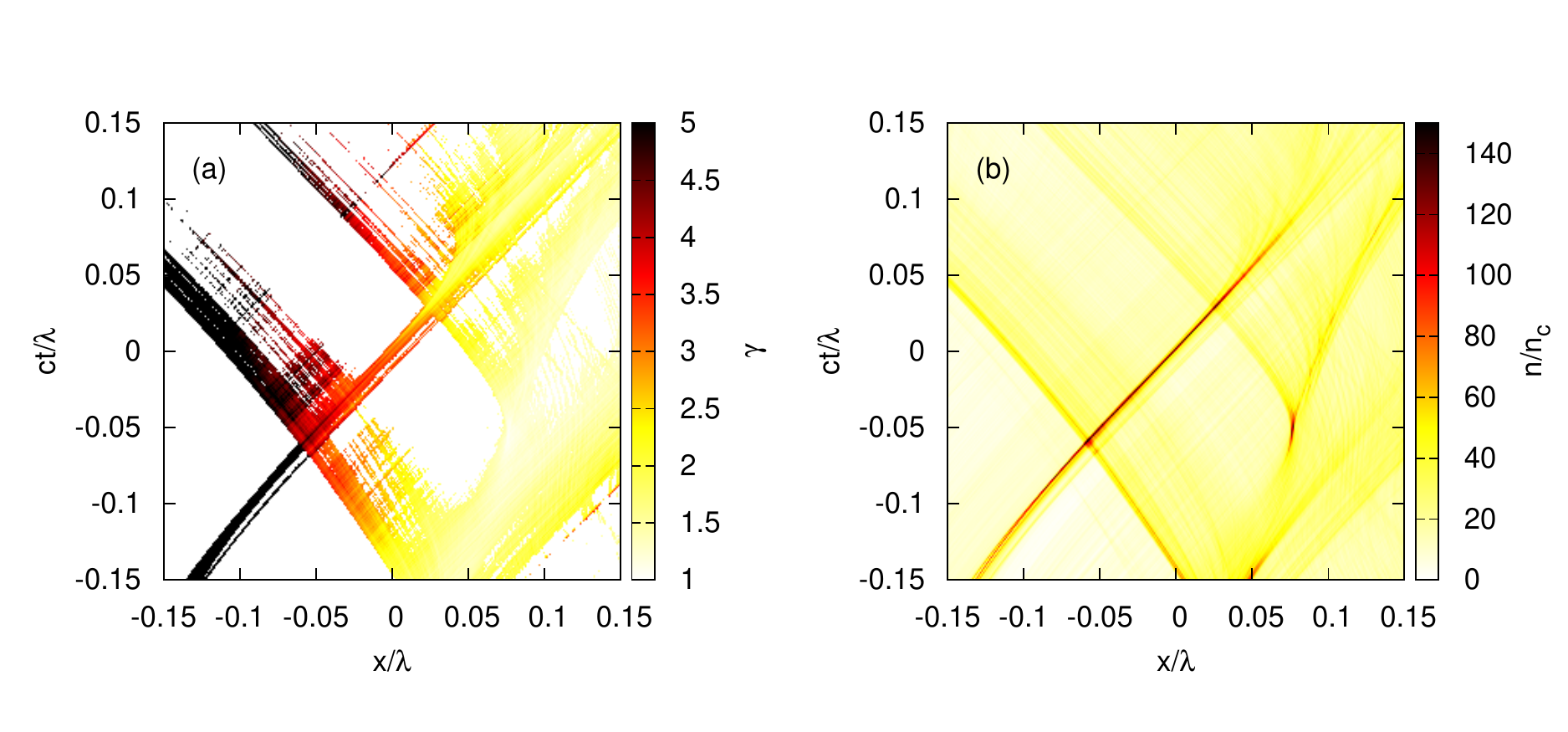} 
	\caption{\small (a) Distribution of the gamma factor in the given electron
		bunch in space time domain. Gamma factor is shown only for the cells
		with density above $20n_{c}$. (b) The electron density distribution
		in space time domain. Simulation parameters are the same as for Fig. \ref{ei}.}
	\label{dens_comp4_2D} 
\end{figure}

\begin{figure}[htbp]
	\centering 
	\includegraphics[width=1\textwidth]{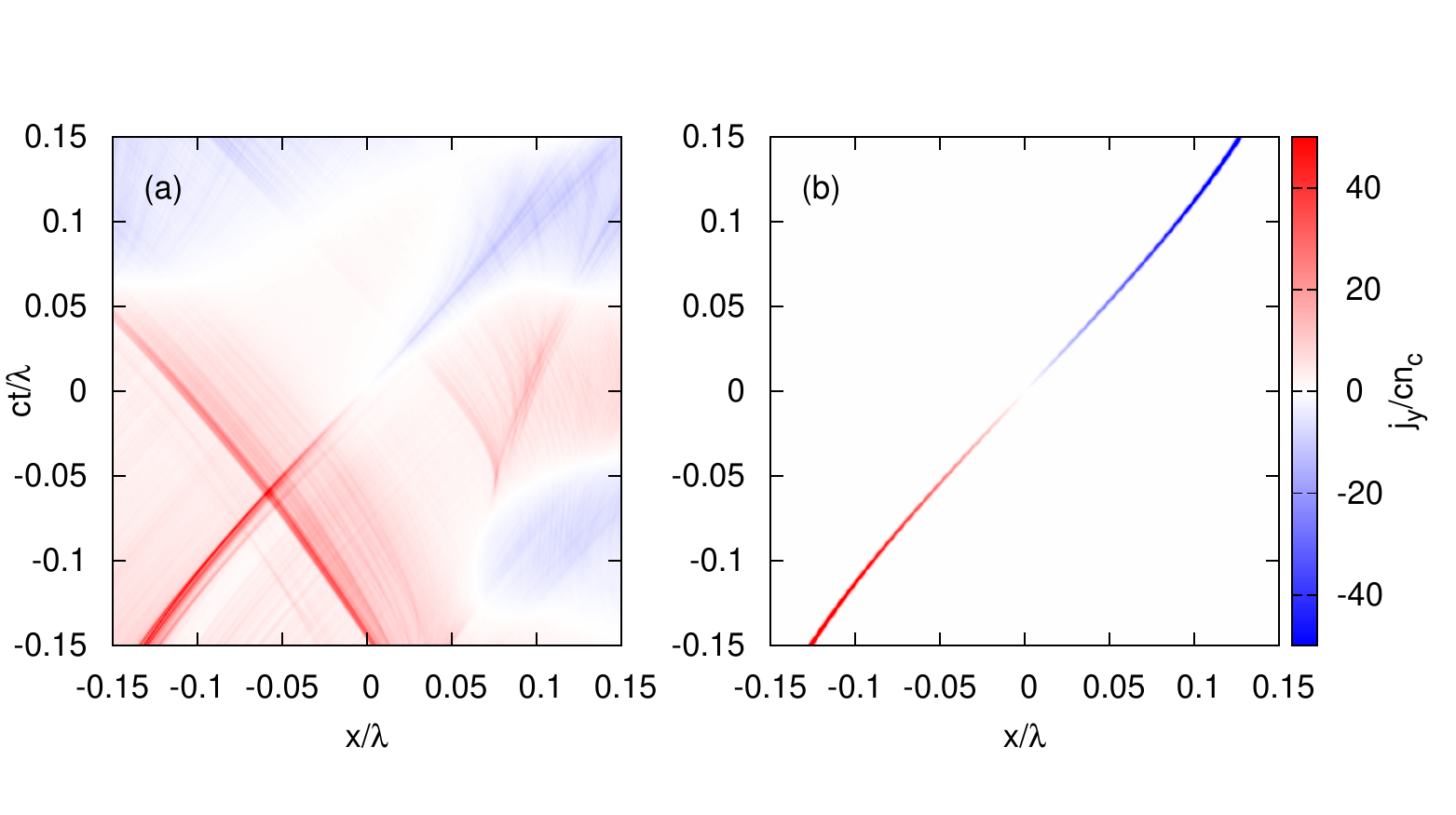} 
	\caption{\small Transverse current density from the simulation (a) and
		calculated analytically (b). In (a) the simulated current density
		near the SPP is illustrated. Simulation parameters are the same as for
		Fig. \ref{ei}. In (b) the analytically calculated current distribution
		is shown. The parameters used in equation (\ref{x_02}) are: $\alpha_{0}=500$,
		$n_{\text{m}}=100$ and $\gamma=3$, while for the shape we used:
		$\tilde{\tilde{\sigma}}=0.0015\lambda$. The velocity $\upsilon$
		in (\ref{x_0}) is obtained from the gamma factor.}
	\label{jy3_2D} 
\end{figure}

\begin{figure}[htbp]
	\centering 
	\includegraphics[width=1\textwidth]{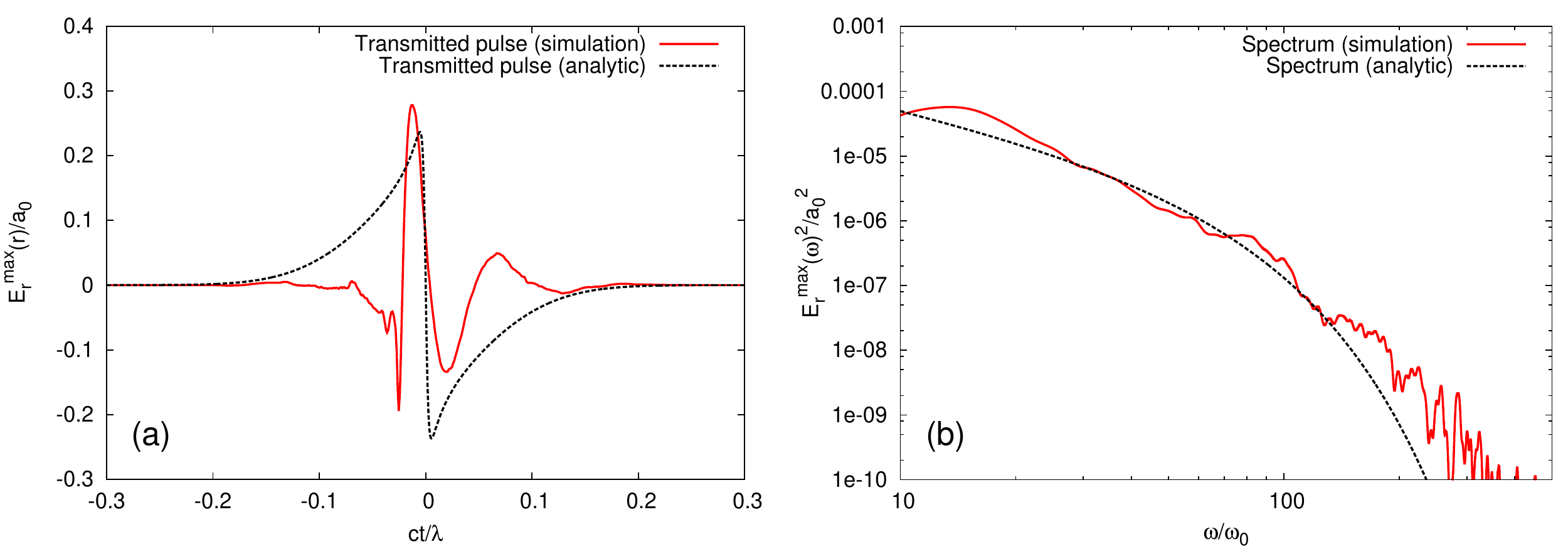} 
		\caption{\small Transmitted radiation obtained from the simulation ((a)
		red) and from the analytical current distribution ((a) black), as well
		as the corresponding spectra in (b). The spectrum from the simulation
		is taken directly from the radiated pulse via FFT, while the other
		one is obtained using the equation (\ref{Ai1}). }
	\label{airy3} 
\end{figure}

We considered only reflected radiation untill now, but the question if the transmitted radiation could also be described
with our model still can be asked. The CSE in transmission has already been obtained
by normal incidence on ultra-thin foils \cite{DRY,DCR}. So we performed
several simulations using the $0.2\lambda$ foil and varying the density
gradient and the density. By choosing $\sigma=0.4$ for the density gradient and
a density of $40n_{c}$, we could obtain a transmitted pulse with a
maximum amplitude that reaches almost $30\%$ of the incident amplitude
(Fig. \ref{ei}). 
This pulse is radiated by the electron nanobunch that is accelerated
in forward direction and reaches a velocity of $\upsilon\approx0.95c$
in the SPP (Fig. \ref{dens_comp4_2D} (b)). 
\begin{figure}[htb]
	\centering 
	\includegraphics[height=7cm]{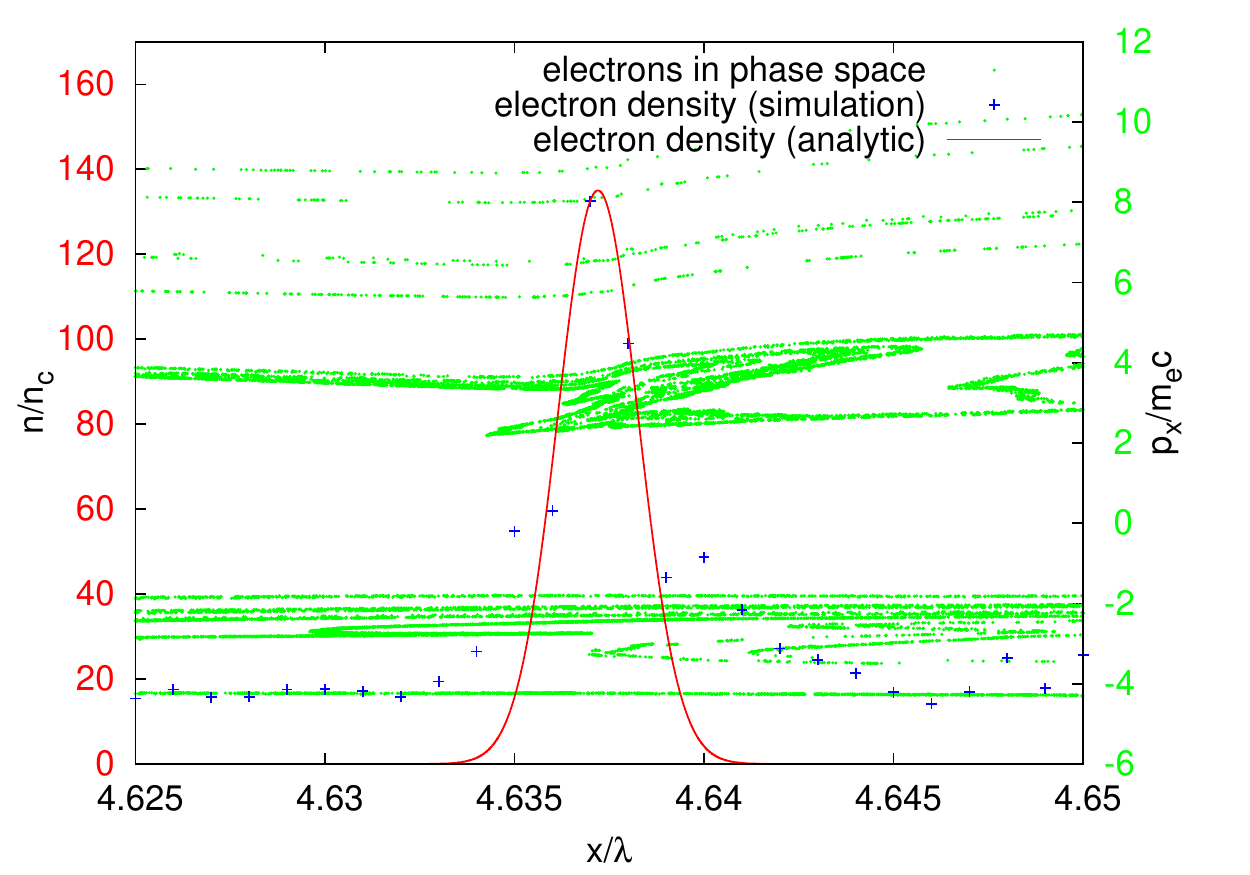} 
	\caption{\small Electron density taken from simulation in SPP (blue) and
		calculated analytically via (\ref{profile_sqrt}) (red), as well as
		electrons in $x$-$p_{x}$-plane (green), Simulation parameters are
		the same comparing to Fig. \ref{ei}, while by analytical description
		we used $\tilde{\tilde{\sigma}}=0.0015\lambda$.}
	\label{fit_dens_gauss} 
\end{figure}
Unfortunately, the electrons are distributed very arbitrarily in phase
space in this case (Fig. \ref{fit_dens_gauss}), so there is no chance
to apply our analytical formulas for the density spike in this case. The
reason of the broad distribution function can be the significant rise of the electron temperature.
Instead, the density profile of the considered electron bunch can
be roughly described with a simple Gaussian 
\begin{equation}
f(x)=e^{-\frac{x^{2}}{\tilde{\tilde{\sigma}}^{2}}}.\label{f2}
\end{equation}
as done in \cite{BP,BP2}. 
In Fig. \ref{jy3_2D} (a) the transverse current distribution of the
given electron bunch is demonstrated. We see that the current changes
its sign in the SPP, so we use equation (\ref{j_teylor2}) here, while
$x_{0}(t)$ changes its sign compared to (\ref{x_02}), 
\begin{equation}
x_{0}(t)\approx\upsilon t-\alpha_{1}\frac{t^{3}}{3}.\label{-x_02}
\end{equation}
Now we can calculate the current distribution of the bunch
that is shown in Fig. \ref{jy3_2D} (b). In order to calculate the
transmitted radiation, we use 
\begin{equation}
E_{tr}(t)=\pi\int j_{\bot}(t+x,x)dx,\label{ei2}
\end{equation}
while the formula for the spectrum is obviously the same as in (\ref{Ai1}).
Since we use a Gaussian function as the shape here, we insert its analytical
Fourier image in (\ref{Ai1}). Subsequently, as in the cases of reflected
radiation, we consider the analytical and numerical transmitted pulses
as well as their spectra (Fig. \ref{airy3}). 
It is not very surprising that the results do not fit exactly, especially
for the low frequency range. This is because the radiation is formed
within the skin layer and has to propagate through whole foil and the
trajectory of the radiating electron nanobunch is not anymore in vacuum,
but encompasses the bulk plasma. 

\newpage
\thispagestyle{empty}

\section{Conclusions}

We were able to obtain two different analytical expressions of the electron density
profile describing the density spikes in two different cases specified
by the electron phase space distribution. First, we presented the
parabolic case, where the phase space distribution can be approximated
by a parabola. In the second case, the electrons in phase space
could be fitted with an exponential function. We called this case
the whip case. A few examples, where the analytical formulas describe
the simulated density quite well have been presented . 
Furthermore, we discussed simulation results of HHG, where we were able to obtain an
amplitude increase in the reflected pulse by a factor of five without
using extremely intense incident laser pulses. This was possible after we
found optimal parameters for the density gradient combined with an optimal
incident angle. Moreover, based on some simple assumptions, we were able
to describe the distribution of transverse current in the vicinity
of the SPP analytically in both cases. The obtained expressions together with
the analytical expressions for electron density give us the possibility
to fit the numerically obtained spectra of the back radiated pulse quite
good.

\section{Acknowledgment}

This work has been supported by DFG TR18 and EU FP7 Eucard-2 projects.

\appendix
\section{Appendix}
\label{ch:appendix}
\subsection{Density profile from phase space distribution containing delta function}
\subsubsection{Parabolic case}
	In chapter \ref{ch:DensProf} we started with the equation
	\begin{equation}
		x(p,t)=x_{0}(t)+\alpha(t)(p-p_{0}(t))^{2},
	\end{equation}
	which locally  describes a curve in phase space.   
	The distribution function of the electrons is given by 
	\begin{align}
	f(x,p,t) & ={\cal C}\delta\left(x-x_{0}(t)-\alpha(t)\left(p-p_{0}(t)\right)^{2}\right),\label{phase_space1}
	\end{align}
	where ${\cal C}$ is a normalization constant. In order to get the
	expression of density we have to perform the integration in momentum
	space 
	\begin{equation}
	n(x,t)=\int dpf(x,p,t).\label{dens_int}
	\end{equation}
	By using well known integration properties of the Dirac delta function
	and doing some algebra we obtain the expression 
	\begin{equation}
	n(x,t)=\frac{{\cal C}}{\sqrt{\alpha(t)\left(x-x_{0}(t)\right)}}.\label{dens1}
	\end{equation}
	Note that this equation makes sense only for $x>x_{0}$. For $x<x_{0}$
	the density has to be zero in this model. This is true because the
	argument of delta function in (\ref{phase_space1}) as the function
	of $p$ vanishes only for $x>x_{0}$. In other words,
	there are no electrons on the left hand side of $x_{0}$. In order to calculate the constant ${\cal C}$,
	we initially write an equation for the number of particles in the interval
	$\Delta x$ by integrating the density from $x_{0}(t)$ to $x_{0}(t)+\Delta x$
	\begin{equation}
	N_{\Delta x}(t)={\cal C}\int_{x_{0}(t)}^{x_{0}(t)+\Delta x}\frac{dx}{\sqrt{\alpha(t)\left(x-x_{0}(t)\right)}}=2{\cal C}\sqrt{\frac{\Delta x}{\alpha(t)}}.\label{N}
	\end{equation}
	We solve the obtained equation for ${\cal C}$ and insert
	it into equation (\ref{dens1}). Finally, we obtain the expression
	for the electron density profile, 
	\begin{equation}
	n(x,t)=\frac{1}{2}\frac{N}{\sqrt{\Delta x\left(x-x_{0}(t)\right)}},
	\end{equation}
	where $N$ is the number of particles contained between $x_{0}(t)$
	and $x_{0}(t)+\Delta x$. Note that the parameter $\alpha$ cancels,
	so it does not affect the density profile. 
\subsubsection{Whip case}
	In chapter \ref{ch:DensProf} we started with the equation
	\begin{equation}
		x_{p}(p)=e^{\alpha p}\label{expApp}.
	\end{equation}
	which locally  describes a curve in phase space.   
	The distribution function of the electrons is given by 	
	\begin{align}
	f(x,p) & ={\cal C}\delta\left(x-x_{p}(p)\right)\label{phase_space2}.
	\end{align}
	Going along the same line as in the previous case, we obtain
	\begin{equation}
	n(x)={\cal C}\int dp~\delta\left(x-x_{p}(p)\right)=\frac{{\cal C}}{\alpha x}\label{dens1/x}.
	\end{equation}
	Obviously equation (\ref{expApp}) can not be applied at the whole interval
	$[0:\Delta x]$ as in the parabolic case since the momentum of the
	electrons is limited by some amount, let say, $p_{\text{cut}}$. Therefore
	the description (\ref{expApp}) is valid only on some interval $[x_{\text{min}}:x_{\text{max}}]$,
	where $x_{\text{min}}=e^{-\alpha p_{\text{cut}}}$. Strictly speaking
	by performing the integration in (\ref{dens1/x}) we have to take
	$-p_{\text{cut}}$ as a lower limit, instead of $-\infty$. 
	This would not change the form of the result but the interval on which
	it is valid, namely for $x>x_{\text{min}}$. Consequently, we integrate
	the expression (\ref{dens1/x}) from $x_{\text{min}}$ to $x_{\text{max}}$
	in order to calculate ${\cal C}$ and get 
	\begin{equation}
	n(x)=\frac{N}{\ln{\left(\frac{x_{\text{max}}}{x_{\text{min}}}\right)}x}.
	\end{equation}
\subsection{Density profile from generalized phase space distribution}
	At first we are going to find an appropriate definition of the function $\delta_a$ with
	\begin{equation}
		\lim_{a\rightarrow0}\delta_{a}(x)=\delta(x).
	\end{equation}
	Gaussian function would fulfill these conditions,
	but if we use it in order to define $\delta_{a}$ we would not be able
	to solve the integral (\ref{dens_int}) analytically. That's why we
	define: 
	\begin{align*}
	g_{a}(x) & \equiv\frac{3}{4a}\left(1-\frac{x^{2}}{a^{2}}\right),\\
	\delta_{a}(x) & \equiv\left\lbrace \begin{aligned} & g_{a}(x)\quad\text{for}\quad x\in[-a,a],\\
	& 0\qquad\text{otherwise.}
	\end{aligned}
	\right.
	\end{align*}
	It is easy to check that with this definition $\delta_{a}(x)$ does
	satisfy the condition (\ref{delta_conv}).

	Let us again calculate the electron density profile for the parabolic
	case with $x(p)=\alpha p^{2}$, which holds if the electron bunch
	moves slowly. We have 
	\[
	n_{a}(x)={\cal C}\int dp~\delta_{a}\left(x-\alpha p^{2})\right).
	\]
	This integration is more complicated as compared to the simple $\delta$-function
	case. We have to be careful with integration boundaries, since $\delta_{a}$
	is a bounded support function. As a result we obtain: 
	\begin{align*}
	n_{a}(x)=\left\lbrace \begin{aligned} & \frac{2{\cal C}}{5a^{3}\sqrt{\alpha}}\left(3a^{2}-2x^{2}+ax\right)\sqrt{x+a}\qquad\qquad\qquad\text{for}\quad x\in[-a,a]\\
	& \frac{2{\cal C}}{5a^{3}\sqrt{\alpha}}\biggr(\left(3a^{2}-2x^{2}\right)\left(\sqrt{x+a}-\sqrt{x-a}\right)\\
	& \qquad\qquad\qquad\quad+ax\left(\sqrt{x+a}+\sqrt{x-a}\right)\biggr)\qquad\text{for}\quad x>a\\
	& \qquad0\qquad\qquad\qquad\qquad\qquad\qquad\qquad\qquad\qquad\text{for}\quad x<-a.
	\end{aligned}
	\right.\\
	\end{align*}
	It is straight forward to show that 
	\begin{align}
	\lim_{a\rightarrow0}n_{a}(x)=n(x)=\left\lbrace \begin{aligned} & \frac{{\cal C}}{\sqrt{\alpha x}}\\
	& 0
	\end{aligned}
	\right.,\label{a->0}\\
	\nonumber 
	\end{align}
	compare with (\ref{dens1}). For the number of particles $N_{a,\Delta x}$
	on interval $[-a:\Delta x]$, that means 
	\begin{equation}
	N_{a,\Delta x}=2{\cal C}\sqrt{\frac{\Delta x}{\alpha}}\quad\text{for}\quad a\ll\Delta x,\label{N_a}
	\end{equation}
	compare with (\ref{N}). The equation (\ref{N_a}) follows directly
	from (\ref{N}) and (\ref{a->0}) for $a\rightarrow0$. Via integration
	of $n_{a}(x)$ on the interval $[-a,\Delta x]$ it can be shown that
	(\ref{N_a}) holds also for $a\ll\Delta x$, which is a more general
	case. Actually we are always able to chose $\Delta x$ in such a way
	that the condition $a\ll\Delta x$ is satisfied and for that case
	we can finally write 
	\begin{align}
		n_{a,\Delta x}(x)=\left\lbrace \begin{aligned} & \frac{N_{a,\Delta x}}{5a^{3}\sqrt{\Delta x}}\left(3a^{2}-2x^{2}+ax\right)\sqrt{x+a}\qquad\qquad\qquad\text{for}\quad x\in[-a,a]\\
		& \frac{N_{a,\Delta x}}{5a^{3}\sqrt{\Delta x}}\biggr(\left(3a^{2}-2x^{2}\right)\left(\sqrt{x+a}-\sqrt{x-a}\right)\\
		& \qquad\qquad\qquad\quad+ax\left(\sqrt{x+a}+\sqrt{x-a}\right)\biggr)\qquad\text{for}\quad x>a\\
		& \qquad0\qquad\qquad\qquad\qquad\qquad\qquad\qquad\qquad\qquad\text{for}\quad x<-a.
		\end{aligned}
		\right.\nonumber \\
	\end{align}
\newpage{}

\end{document}